\documentclass[lettersize,journal]{IEEEtran}

\usepackage{amsmath,amsfonts}
\usepackage{algorithmic}
\usepackage{algorithm}
\usepackage{array}
\usepackage{graphicx}
\usepackage{textcomp}
\usepackage{stfloats}
\usepackage{url}
\usepackage{verbatim}
\usepackage{multirow}
\usepackage{hyperref}
\usepackage{cleveref}
\usepackage{booktabs}
\usepackage[table]{xcolor}
\usepackage{makecell}
\usepackage{pifont}
\usepackage{cite}
\usepackage[hypcap=false]{caption}
\usepackage[caption=false,font=normalsize,labelfont=sf,textfont=sf]{subfig}
\usepackage{cleveref}
\usepackage{cuted}     
\usepackage{caption}   
\usepackage{etoolbox}  
\usepackage{etoolbox}
\makeatletter
\patchcmd{\@maketitle}
  {\vskip 1em}
  {\vskip 0.5em}
  {}{}
\patchcmd{\@maketitle}
  {\normalfont\normalsize\lineskip 0.5em}
  {\normalfont\normalsize\lineskip 0.25em}
  {}{}
\patchcmd{\@maketitle}
  {\vskip 0.5em}
  {\vskip 0.25em}
  {}{}
\makeatother

\newcommand{\etal}{et al.}
\title{\textbf{Niagara}: Normal-Integrated Geometric Affine Field for Scene Reconstruction from a Single View}

\author{Xianzu Wu$^{*}$,~\IEEEmembership{} 
        Zhenxin Ai$^{*}$,~\IEEEmembership{} 
        Harry Yang,~\IEEEmembership{} 
        Sernam Lim,~\IEEEmembership{} 
        Jun Liu,~\IEEEmembership{Senior Member,~IEEE,} 
        Huan Wang$^{\dagger}$,~\IEEEmembership{Member,~IEEE}
\thanks{$^{*}$Equal contribution.  $^{\dagger}$Corresponding author}
\thanks{Manuscript received April 19, 2021; revised August 16, 2021.}
}
\begin{document}

\maketitle
\thispagestyle{empty}
\pagestyle{empty}

\begin{strip}
    \centering
    \includegraphics[width=0.96\textwidth, height=5.0cm, trim=0.2cm 0cm 0cm 0cm, clip]{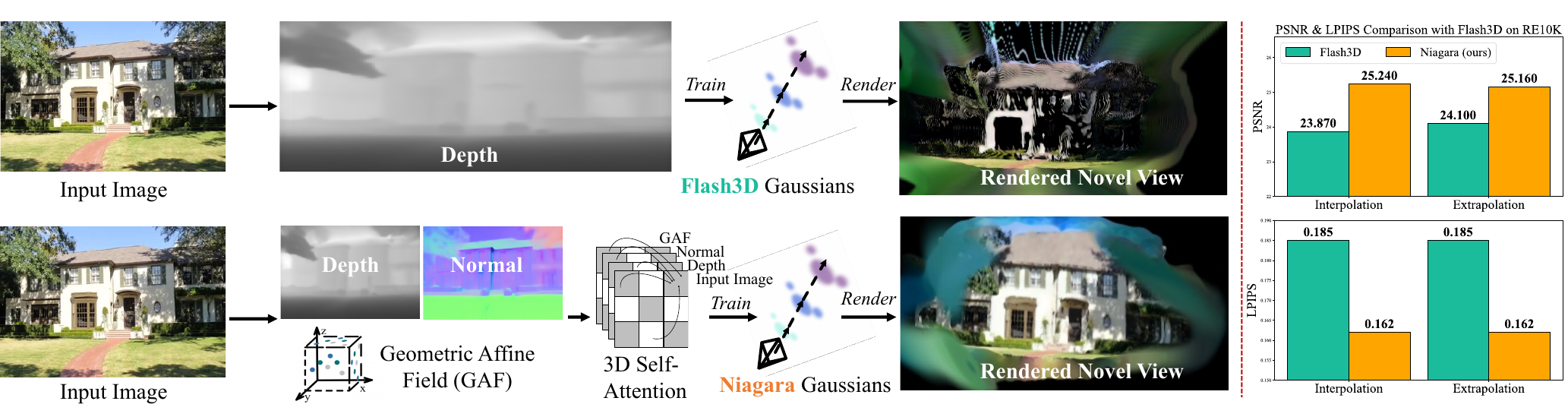}
    \captionof{figure}{
        \textit{Left:} This paper presents \textit{Niagara}, a new 3D scene reconstruction method from a single view.
        Unlike Flash3D~\cite{szymanowicz2024flash3d}, which uses only depth maps, Niagara integrates surface normals and a geometric affine field (GAF) with a 3D self-attention module to better recover scene geometry.
        \textit{Right:} Our method significantly improves novel view synthesis quality, achieving new SoTA results in outdoor scenes.
    }
    \label{fig:teaser}
\end{strip}


\begin{abstract}

Recent advances in \textit{single-view} 3D scene reconstruction have highlighted the challenges in capturing fine geometric details and ensuring structural consistency, particularly in high-fidelity outdoor scene modeling.
This paper presents \textbf{Niagara}, a new \textbf{single-view} 3D scene reconstruction framework 
%
%
that can faithfully reconstruct challenging outdoor scenes from a single input image for the first time.
Our approach integrates monocular depth and normal estimation as input, 
which substantially improves its ability to capture fine details, mitigating common issues like geometric detail loss and deformation.
Additionally,  we introduce a geometric affine field (GAF) and 3D self-attention as geometry-constraint, which combines the structural properties of explicit geometry with the adaptability of implicit feature fields, striking a balance between efficient rendering and high-fidelity reconstruction.
Our framework finally proposes a specialized encoder-decoder architecture, where a depth-based 3D Gaussian decoder is proposed to predict 3D Gaussian parameters, which can be used for novel view synthesis.

Extensive results and analyses suggest that our Niagara surpasses prior SoTA approaches such as Flash3D in both single-view and dual-view settings, significantly enhancing the geometric accuracy and visual fidelity, especially in outdoor scenes.

\end{abstract}

\section{Introduction}\label{sec:intro}
3D scene reconstruction from images has long been a fundamental challenge in the field of computer vision \cite{seitz2006comparison,khoshelham2012accuracy,wu2016learning,choy20163d,mescheder2019occupancy,sitzmann2020implicit}, with extensive applications in various domains such as autonomous driving, drone surveying, game development, virtual reality, and building information modeling~\cite{teichmann2018multinet,nex2014uav,li2019stereo,zingoni2015real,izadi2011kinectfusion,mossel2016streaming}. 
Traditional methods predominantly rely on Multi-View Stereo (MVS)~\cite{furukawa2015multi,hiep2009towards}, which estimates depth maps from multiple images and integrates them to create a comprehensive 3D model  \cite{furukawa2015multi,galliani2015massively,schonberger2016pixelwise,chen2025surfacecontinuous}. 
Recent advances, such as 3D Gaussian splatting \cite{kerbl20233d}, neural radiance fields \cite{mildenhall2021nerf,muller2022instant} or light fields~\cite{bemana2020x,sitzmann2021light,wang2022r2l,cao2023real}, and their derivative works \cite{barron2021mip,barron2022mip}, have effectively addressed previous challenges associated with inconsistent regions, including occlusions, specular reflections, transparent objects, and low-texture surfaces in multi-view scenes.

\begin{figure}
\centering
\begin{tabular}{@{}c@{\hspace{0.001\linewidth}}c@{\hspace{0.001\linewidth}}c@{\hspace{0.001\linewidth}}c@{\hspace{0.001\linewidth}}c@{}}
\includegraphics[width=0.194\linewidth]{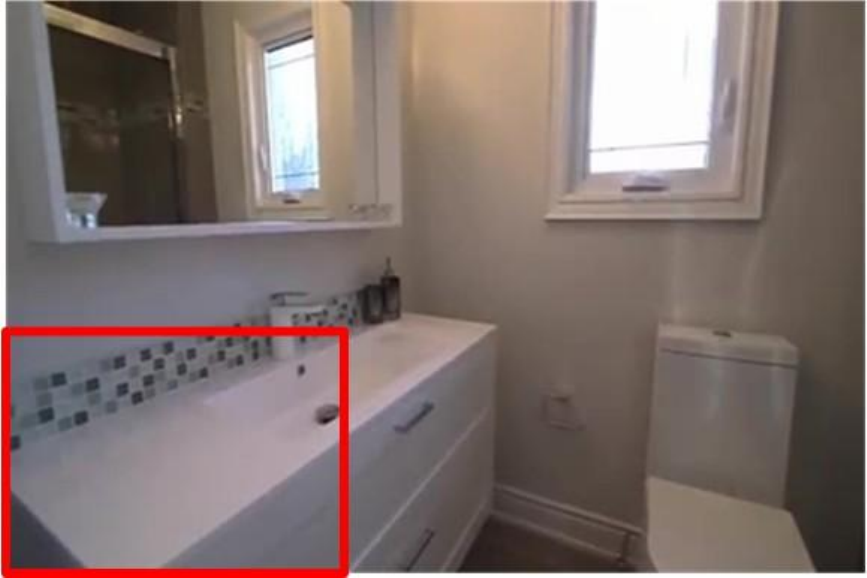} & 
\includegraphics[width=0.194\linewidth]{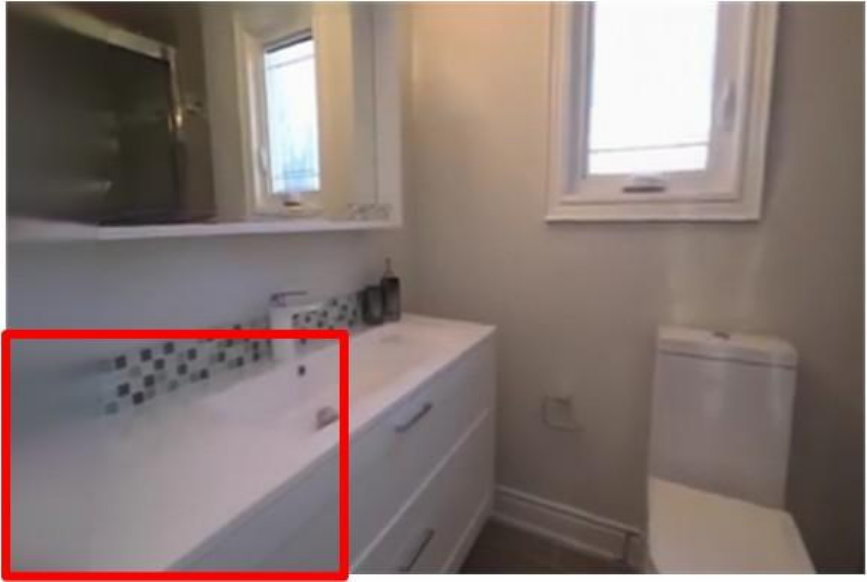} &
\includegraphics[width=0.194\linewidth]{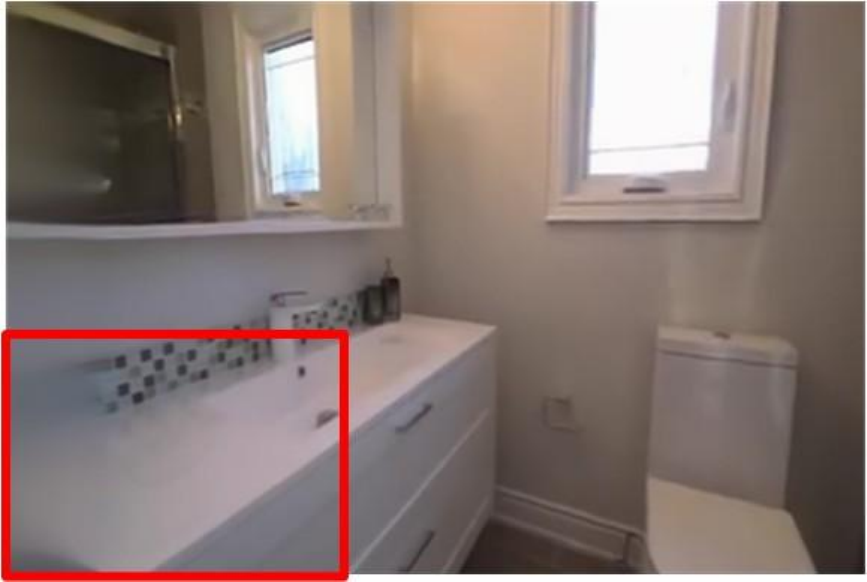} &
\includegraphics[width=0.194\linewidth]{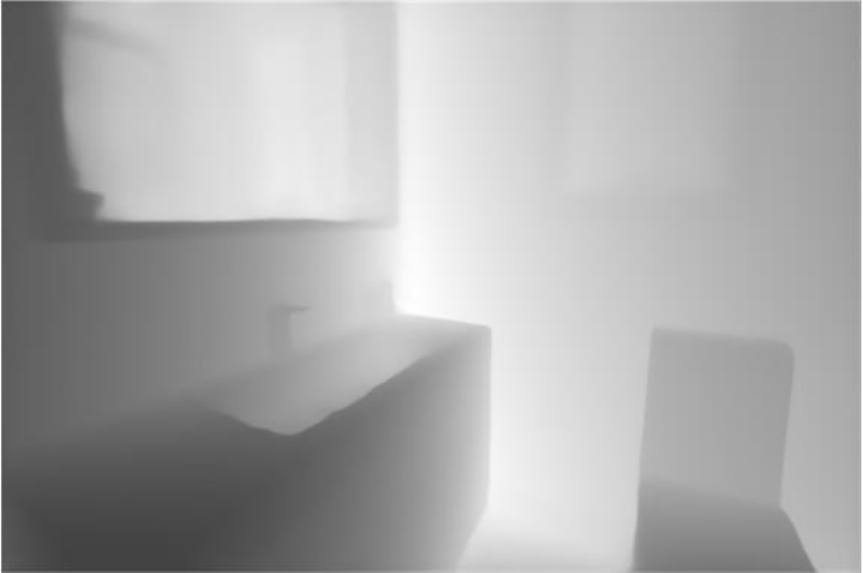} &
\includegraphics[width=0.194\linewidth]{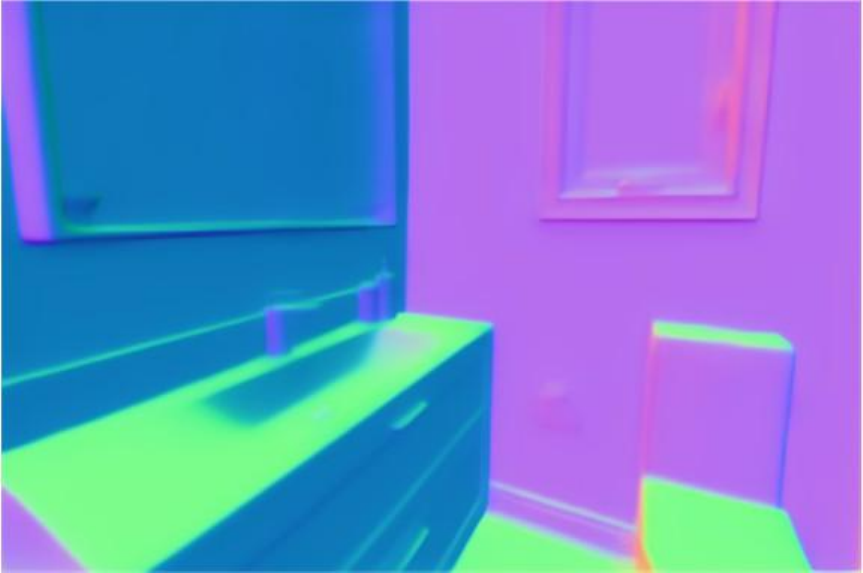} \\
\includegraphics[width=0.194\linewidth]{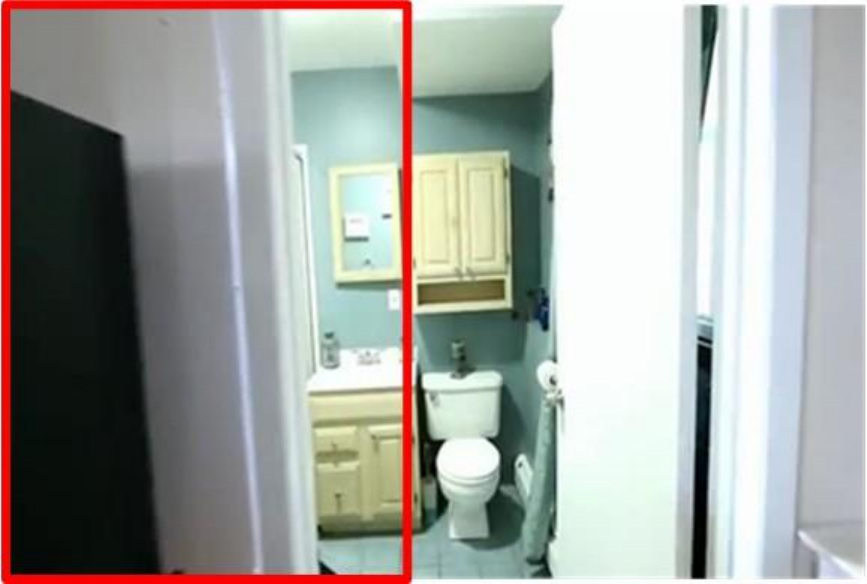} & 
\includegraphics[width=0.194\linewidth]{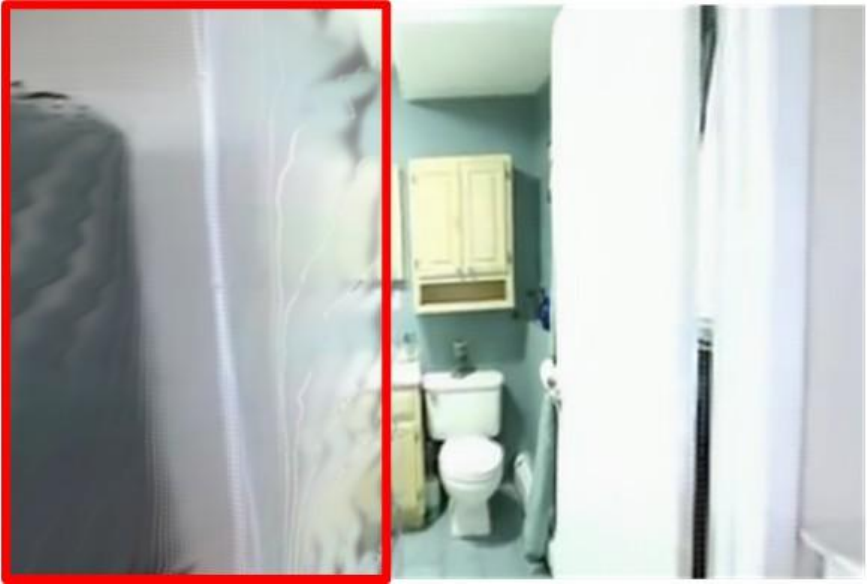} &
\includegraphics[width=0.194\linewidth]{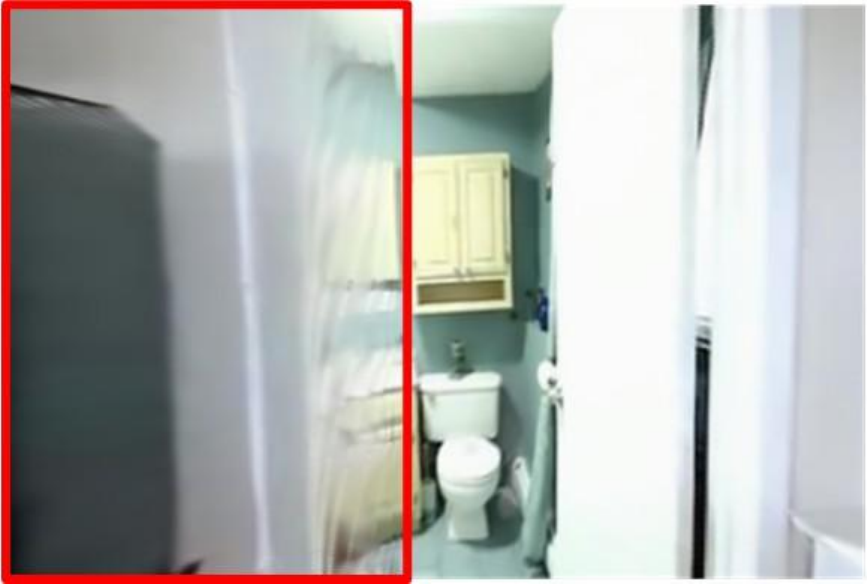} &
\includegraphics[width=0.194\linewidth]{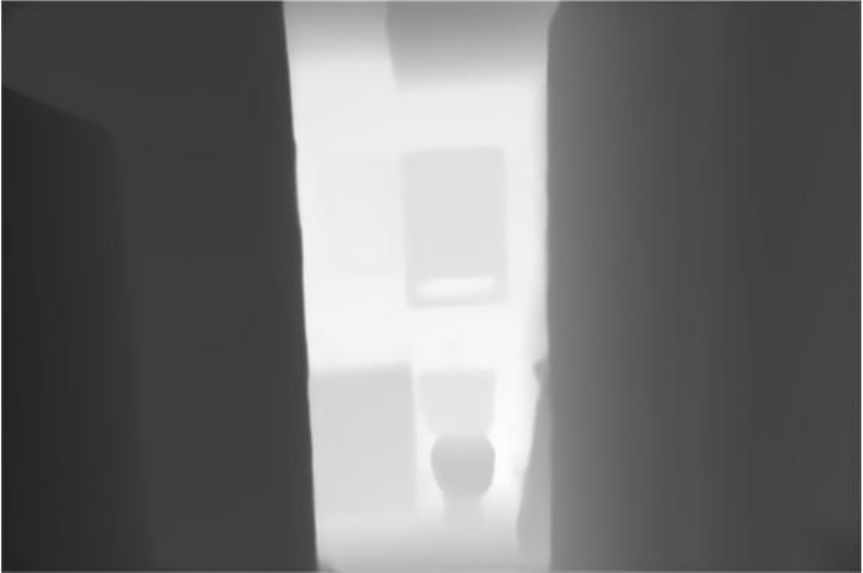} &
\includegraphics[width=0.194\linewidth]{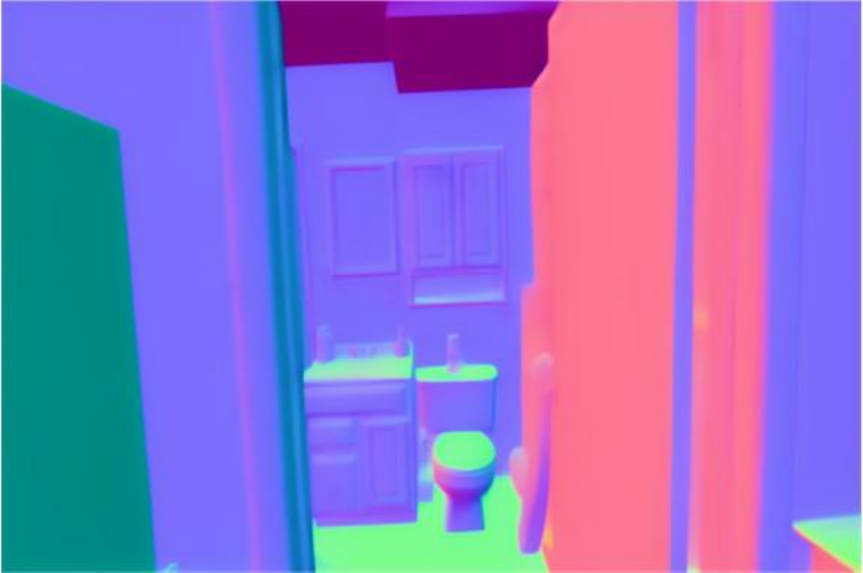} \\
\includegraphics[width=0.194\linewidth]{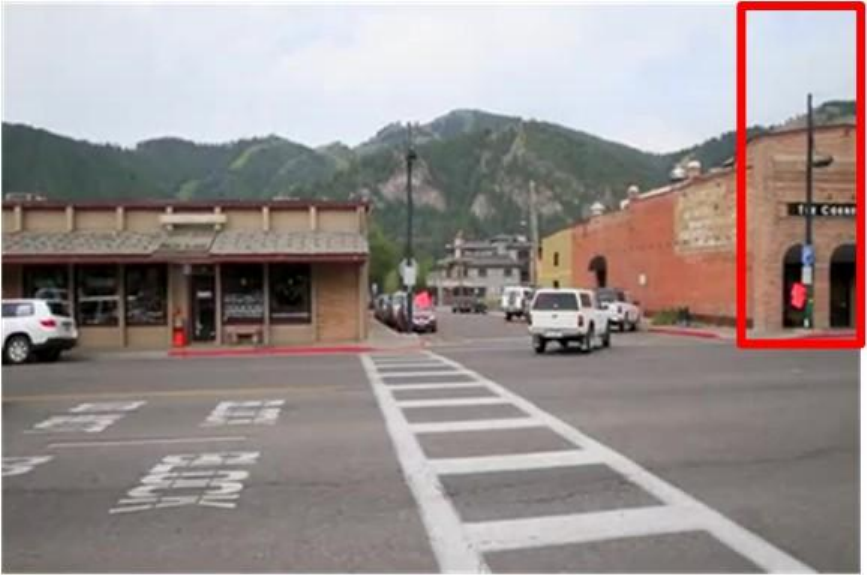} & 
\includegraphics[width=0.194\linewidth]{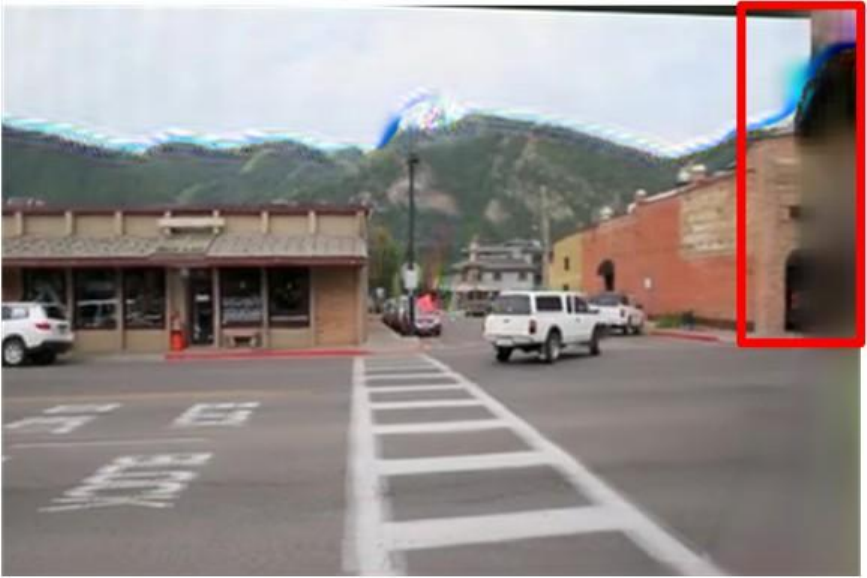} &
\includegraphics[width=0.194\linewidth]{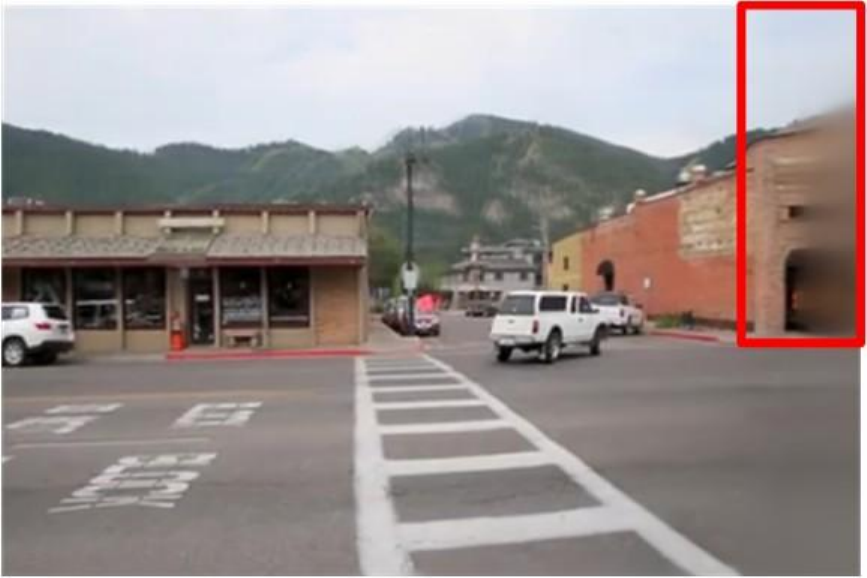} &
\includegraphics[width=0.194\linewidth]{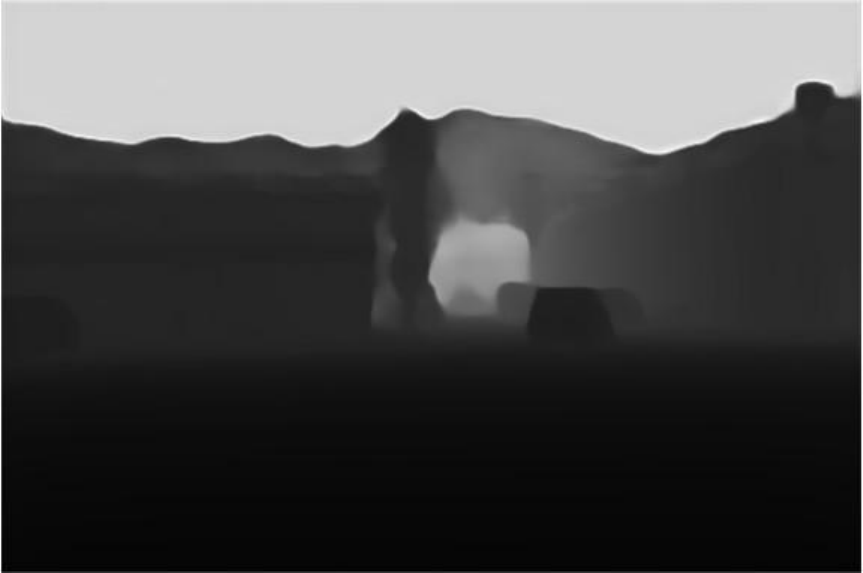} &
\includegraphics[width=0.194\linewidth]{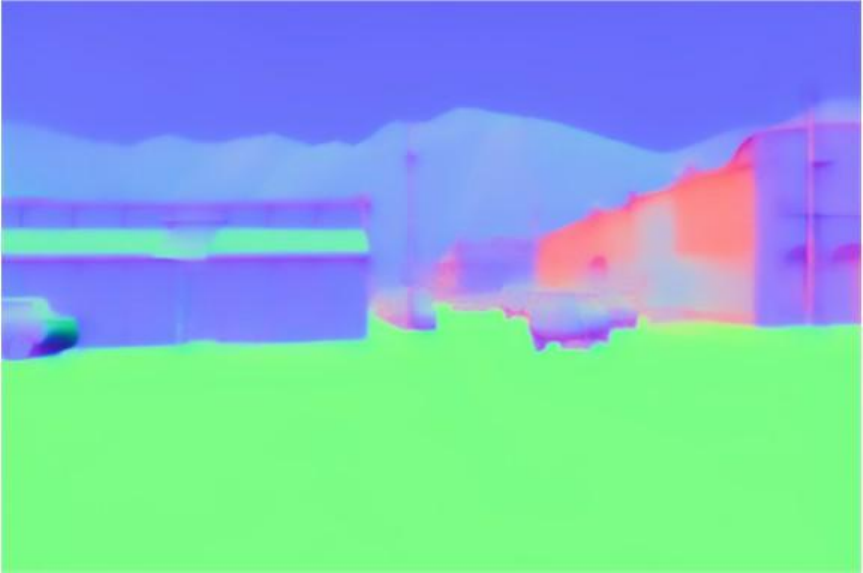} \\
\includegraphics[width=0.194\linewidth]{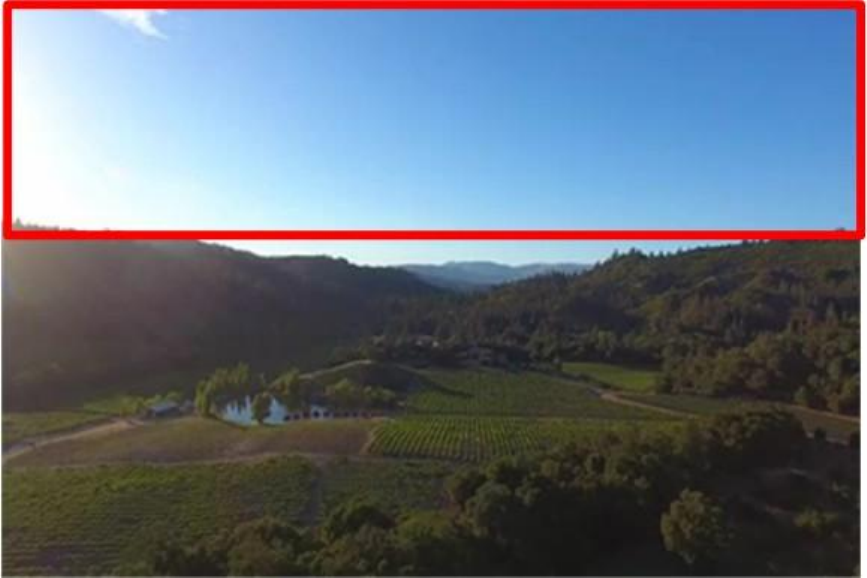} & 
\includegraphics[width=0.194\linewidth]{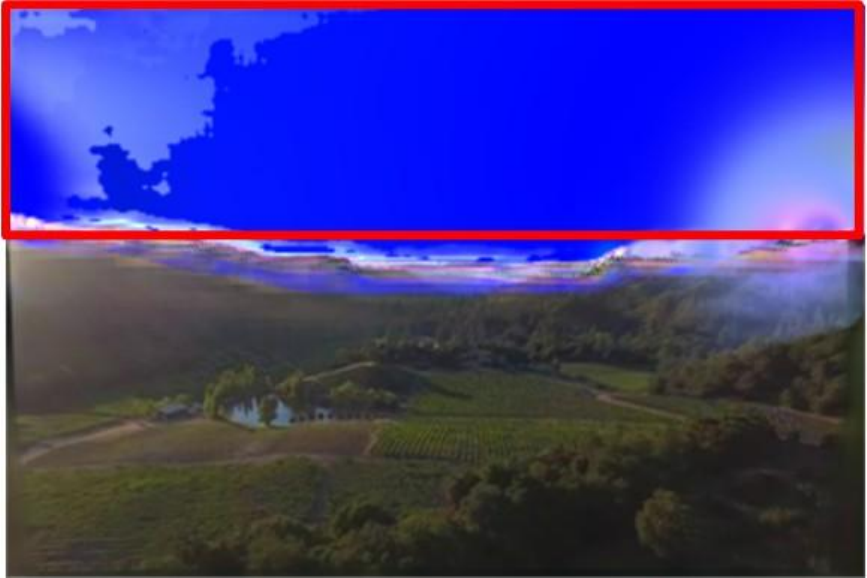} &
\includegraphics[width=0.194\linewidth]{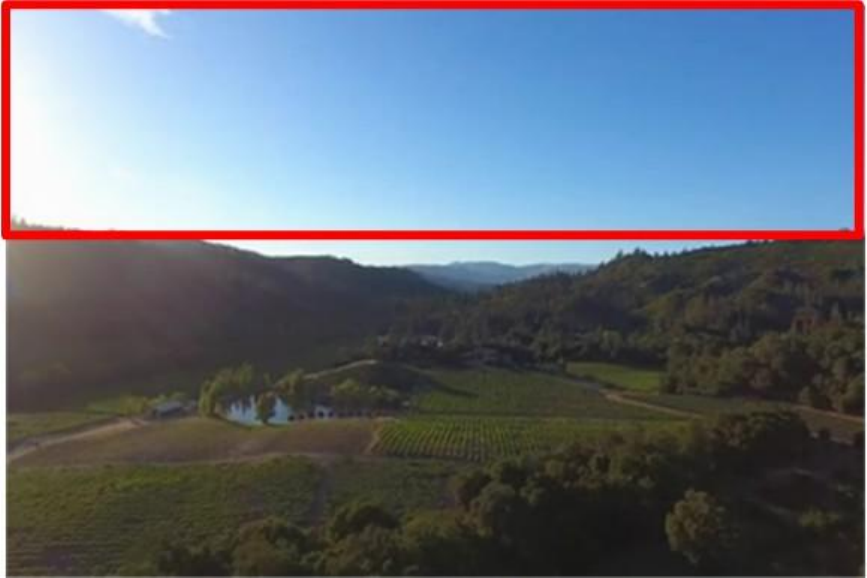} &
\includegraphics[width=0.194\linewidth]{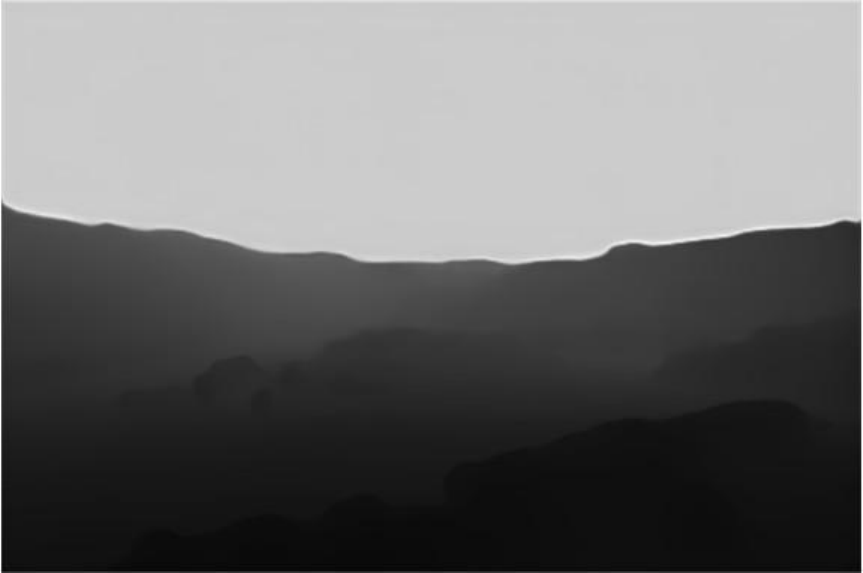} &
\includegraphics[width=0.194\linewidth]{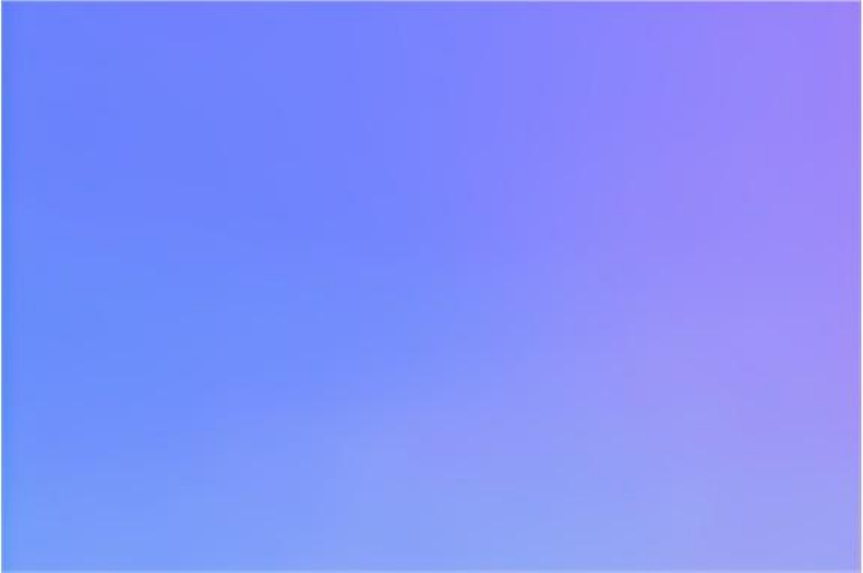} \\
\includegraphics[width=0.194\linewidth]{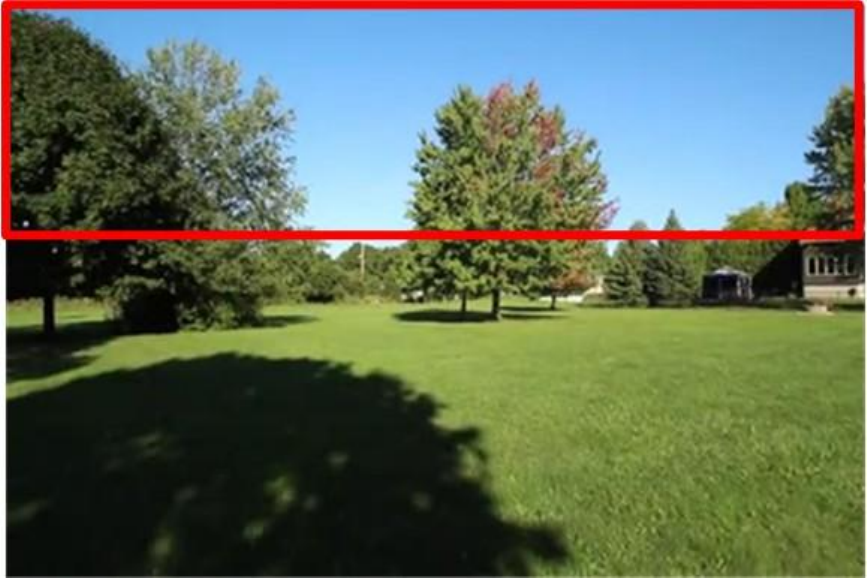} & 
\includegraphics[width=0.194\linewidth]{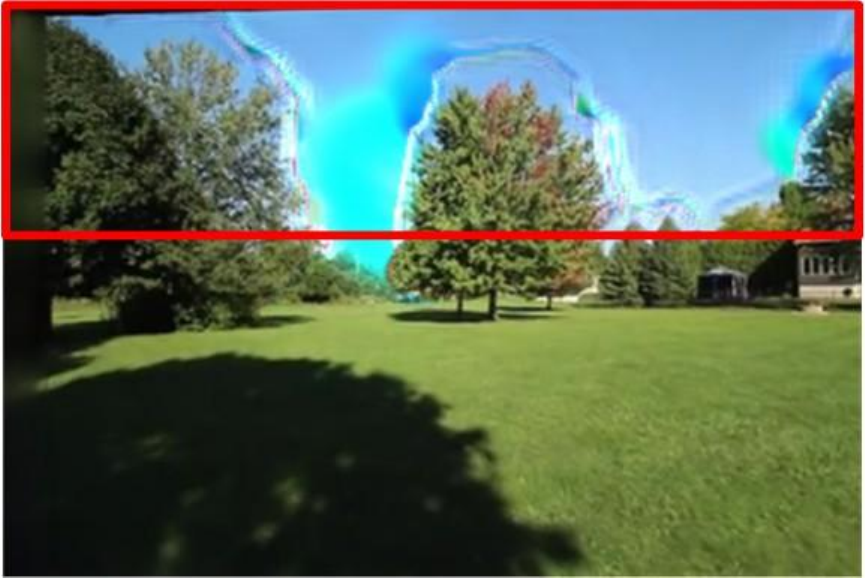} &
\includegraphics[width=0.194\linewidth]{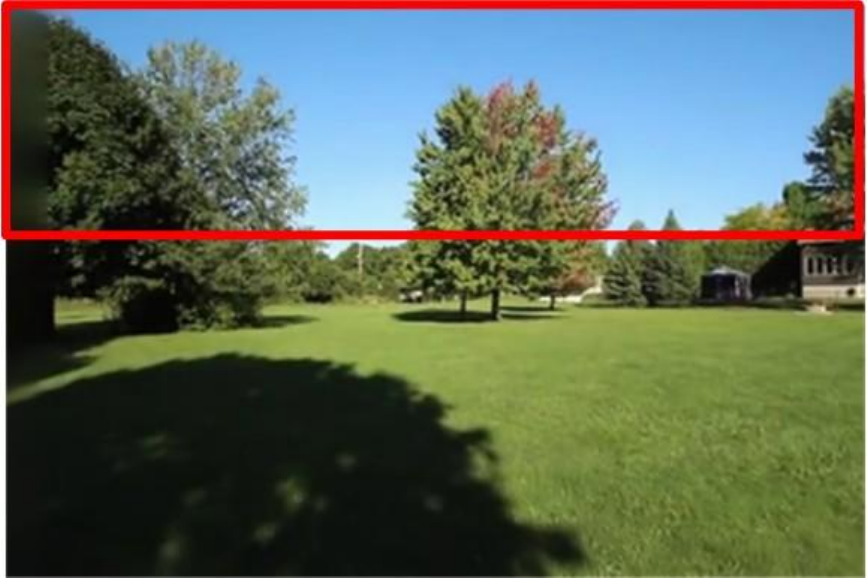} &
\includegraphics[width=0.194\linewidth]{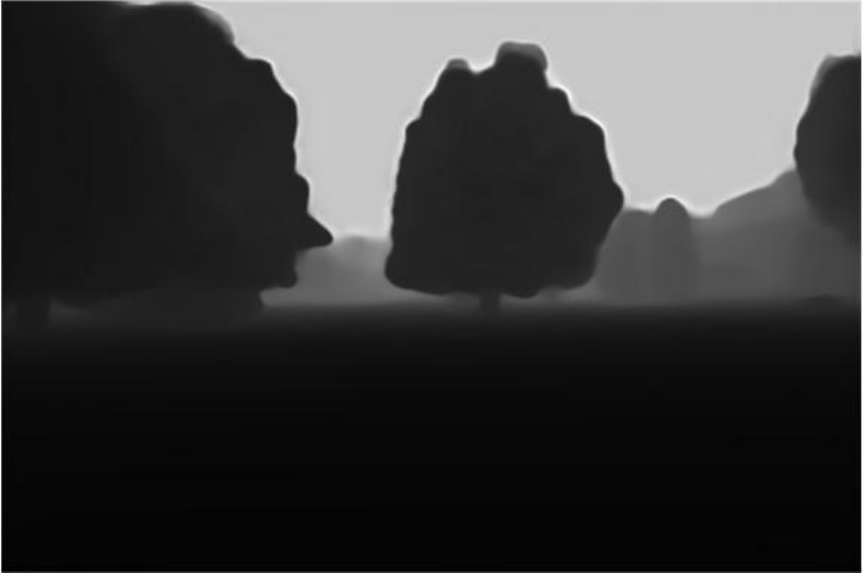} &
\includegraphics[width=0.194\linewidth]{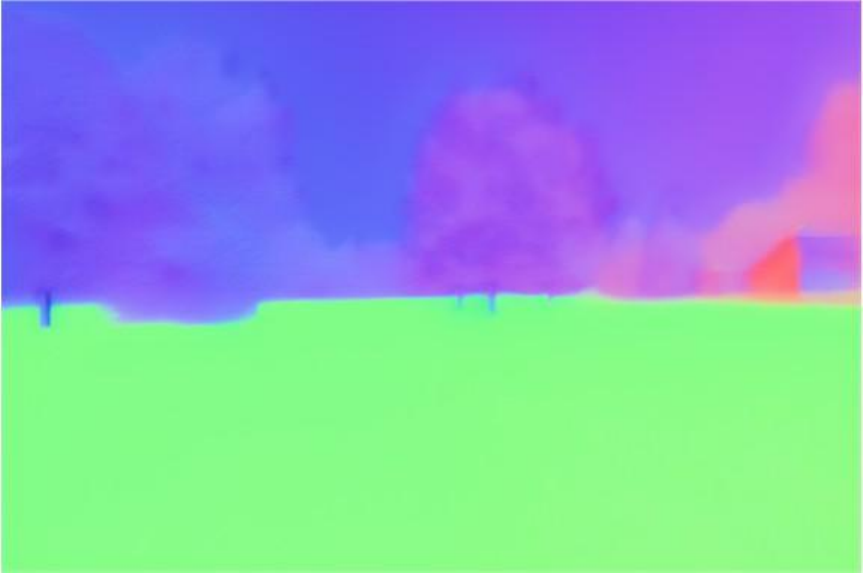} \\
\small (a) GT & \small (b) Flash3D & \small (c) Ours & \small (d) Depth & \small (e) Normal

\end{tabular}
\caption{\textbf{Motivation illustrations.} Flash3D faces geometric blurring and color distortion issues due to Gaussian interpolation errors and insufficient Gaussian representation solely from depth images. To resolve this, we incorporate normal images into our framework, significantly improving reconstruction results.}
\label{fig:Motivation}

\end{figure}

However, \textit{monocular} reconstruction faces significant challenges due to the absence of inherent depth information \cite{saxena20083,li2015depth,gui2024depthfm,wofk2019fastdepth,yang2024depth} and the limitations associated with single-viewpoint data \cite{zhou2018stereo}. Recently, Stanislaw~\etal.~introduced Flash3D~\cite{szymanowicz2024flash3d}, the \textit{first} monocular method, to our best knowledge, that realizes \textit{single-view} 3D reconstruction. Flash3D integrates depth estimation \cite{buxton1983monocular,shao2023nddepth,yang2024depth,birchfield1999depth,li2015depth} with a feedforward network transformer architecture, resulting in notable advances in reconstructing complex scenes without the need for multiview inputs. 

Although Flash3D \cite{szymanowicz2024flash3d} has made significant strides in single-view 3D scene reconstruction, there is still notable room for improvement. In \autoref{fig:Motivation}, we illustrate several examples in which Flash3D did not perform so well. The first two rows show Flash3D renders \textit{blurred} results for the corners and door edges with abrupt geometric changes. This blurring is caused by insufficient depth interpolation, resulting in a loss of geometric fidelity. 
Inadequate point cloud \cite{zhang2023frequency,wu2024fsc,tong2024ywnwa} sampling exacerbates interpolation errors \cite{rusinkiewicz2000qsplat}, particularly in detail-rich regions such as the edges or boundaries of physical structures and the sky. Color distortion and overflow artifacts (see \autoref{fig:Motivation}, Row 4, 5) occur due to insufficient Gaussian representation.

To address these shortcomings, this paper proposes a new \underline{n}ormal-\underline{i}ntegr\underline{a}ted \underline{g}eometric \underline{a}ffine field for 3D scene \underline{r}econstruction from \underline{a} single view, abbreviated as \textit{Niagara}. Niagara aims to improve geometric accuracy while preserving fine details, as shown in \autoref{fig:teaser}.
Specifically, we integrate both depth and normal information \cite{fan2020sne,hu2024metric3d,ye2024stablenormal}, improving depth cues and allowing the model to capture finer details. Furthermore, we introduce a  \textit{geometric affine field} (GAF) with 3D self-attention as geometric constraint, which enriches the geometric details in single-view image reconstruction and enhances the sensitivity of the model to geometric boundaries \cite{wimbauer2023behind,chang2015shapenet,bhoi2019monocular}.
The output of GAF is then used to learn the parameters of 3D Gaussian splatting (3D-GS) ~\cite{huang2018deepmvs,chen2025mvsplat,szymanowicz2024flash3d,hu2024metric3d,long2024wonder3d,zhang2025toy}, which can be used to render novel views during testing.
Experimental results show that Niagara excels in a variety of challenging scenes, offering notable advantages in terms of geometric accuracy and preservation of detail when compared to existing methods (a quick demonstration is shown in \autoref{fig:Motivation} - our method outperforms Flash3D, the prior SoTA in this line).

Our contributions can be summarized as follows:
\begin{itemize}
    \item This paper proposes Niagara, the \textit{first} effective \textit{single-view} 3D scene reconstruction framework that addresses key challenges for complex outdoor scenes, by integrating surface normals for improved global features.
   
    \item Niagara introduces several novel modules for accurate scene representation and learning: a geometric affine field and 3D self-attention for refining local geometry, and a depth-based Gaussian decoder for novel view rendering.
    \item Experiments demonstrate Niagara achieves state-of-the-art performance on the single-view 3D reconstruction benchmark, surpassing existing methods by nearly 1 dB PSNR, especially in challenging outdoor scenes.
\end{itemize}

\section{Related Work}
\label{sec:formatting}

\begin{figure*}
    \centering
    \includegraphics[width=1\linewidth]{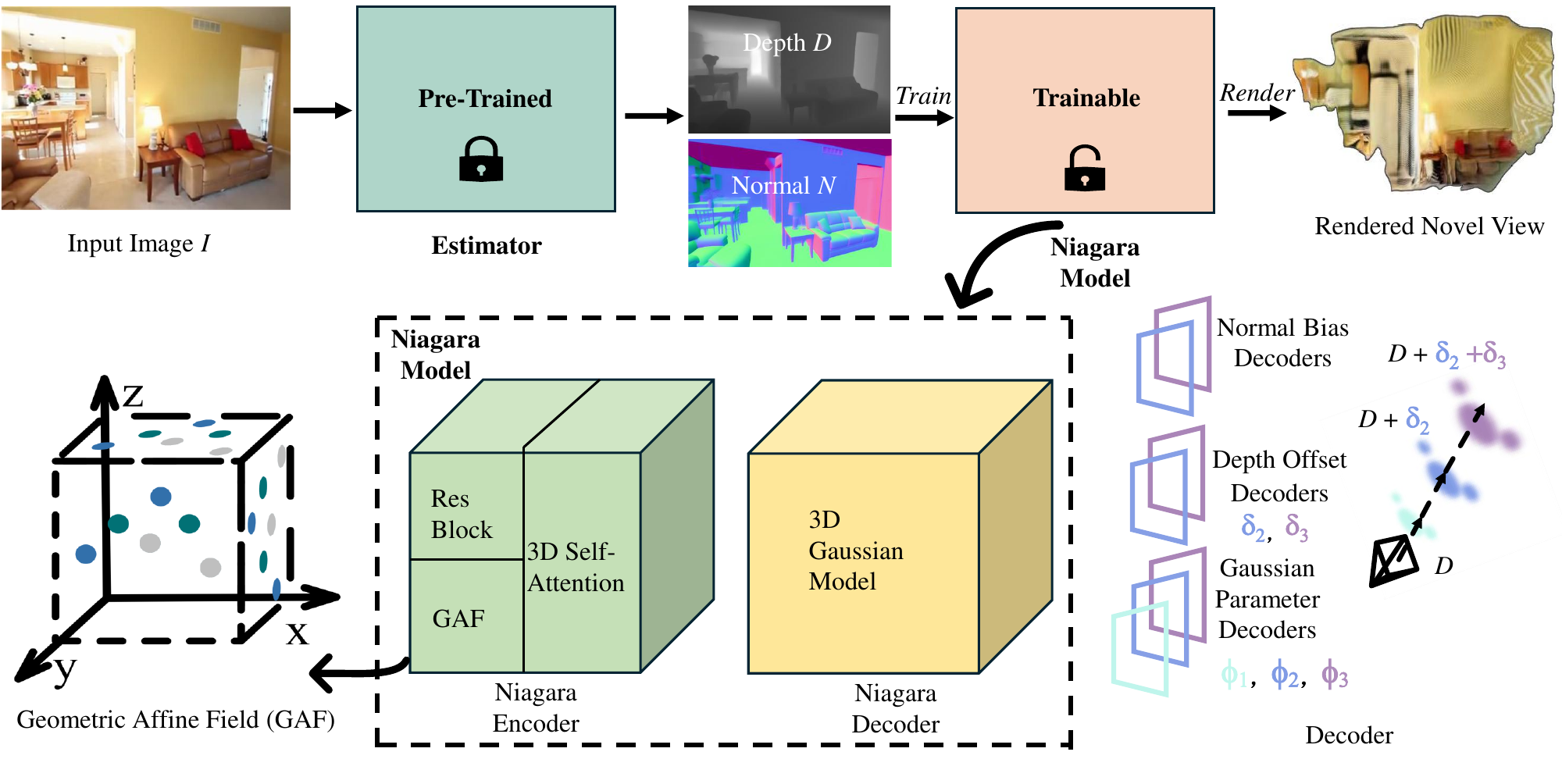}
    \caption{\textbf{Overview of Niagara. } First, two frozen pre-trained networks \cite{piccinelli2024unidepth,ye2024stablenormal} simultaneously estimate the metric depth map \( D \) and the normal map \( N \) of the input image \( I \). 
    \textbf{GAF} (Geometric Affine Field) module and the Res block in the encoder-decoder is ResNet50 based and combines with a 3D self-attention module. Niagara Decoder is similar to that of Flash3D~\cite{szymanowicz2024flash3d}. Niagara Decoder predicts shape and appearance parameters \( \hat{P} \) for \( K \) layers of Gaussian distributions at each pixel \( u \). By adding the predicted positive depth offsets \( \delta_i \) to the initial monocular depth \( D \), the depth for each Gaussian layer is obtained. At the same time, the estimated normal information is used to compute the mean vectors of each Gaussian layer. This approach ensures that the Gaussian slices are ordered by depth, effectively modeling occluded and unobserved surfaces and increasing the accuracy of 3D reconstruction from a single image. }
     \label{fig:Overview}
     \vspace{-5mm}
\end{figure*}

\textbf{\textit{Single-View} 3D Scene Reconstruction.} \textit{Single-view} 3D scene reconstruction typically involves per-pixel depth estimation \cite{duzceker2021deepvideomvs}. 
However, independent pixel-wise depth computation can result in artifacts \cite{duzceker2021deepvideomvs}, compromising global geometric continuity and concave boundary consistency, particularly when faced with complex shapes.

Neural networks are increasingly used for single-view reconstruction. 
Wiles \etal.~\cite{wiles2020synsin} introduce a novel end-to-end view synthesis method that leverages depth-guided 3D point clouds in combination with neural rendering. However, this method faces challenges, such as noticeable distortions and blurring, particularly when subjected to significant changes in viewpoint.

Li \etal.~\cite{li2021neural} combine Neural Radiance Fields (NeRF) \cite{mildenhall2021nerf} with Multi-Plane Images (MPI) \cite{dai2020neural} for novel view synthesis, effectively capturing depth variations. Nonetheless, this method may struggle in complex scenes characterized by substantial depth or lighting variations.

Recently, Wofk \etal.~\cite{wofk2019fastdepth} developed a fast and efficient reconstruction technique that leverages convolutional neural networks in conjunction with spatial transformations. Despite its speed, the approach may sacrifice fine details and depth consistency, especially in intricate scenes with significant changes in viewpoint. 

\vspace{0.3em}
\noindent\textbf{Monocular Depth and Normal Estimation.} Our approach builds on monocular depth estimation, which predicts pixel-wise depth in images. Recent methods in this line have built upon deep neural networks and shown remarkable advances~\cite{szymanowicz2024flash3d,chen2025mvsplat,charatan2024pixelsplat,yu2025vfmdepth,feng2025efficient}, driven by large training datasets \cite{zhou2018stereo,silberman2012indoor}. Due to varying depth distributions under different RGB values, some methods discretize depth as a classification task to enhance performance~\cite{geiger2013vision}.

Depth models in 3D reconstruction face two main challenges: adapting to diverse scenes and accurately predicting metric information under varying camera settings. Bhat \etal. \cite{bhat2021adabins} introduces global depth distributions with conditional scene-based processing. Yin \etal. \cite{yin2019enforcing} scale inputs and outputs to standard space, remapping depth by focal length. Facil \etal. \cite{facil2019cam} employ camera intrinsic parameters as inputs to improve depth estimation, while Piccinelli \etal. \cite{piccinelli2024unidepth} leverages self-supervised learning with variational inference for camera-specific embeddings. Gui \etal. \cite{gui2024depthfm} develops a fast monocular depth estimation method using flow matching to enhance efficiency and accuracy.

Surface normals are less prone to metric ambiguity, better capture geometric shapes, and are crucial in 3D reconstruction tasks. However, estimating fine details and avoiding directional bias remains challenging, especially in unstructured scenes. Bae \etal. \cite{bae2024rethinking} utilize pixel ray directions but face difficulties in complex scenes. Long \etal.~\cite{long2024wonder3d} model joint color-normal distribution for consistency, while Ye \etal.~\cite{ye2024stablenormal} refines initial estimates with semantic-guided diffusion to enhance sharpness and reduce randomness. 
In this work, we choose to employ pre-trained models to generate the depth~\cite{piccinelli2024unidepth} and normal maps~\cite{ye2024stablenormal}.
\section{Proposed Method}

Given an input RGB image $I \in \mathbb{R}^{3 \times H \times W}$ that captures a scene, our goal is to train a neural network $\Phi$ to take $I$ as input and generate a scene representation $G = \Phi(I)$, encapsulating both the 3D geometry and the photometric properties of the scene described by~$I$. $G$ will be used to render novel views during testing.
Next, \Cref{sec:background} outlines the basic concepts and framework on which we build. \Cref{sec:Baseline} formally introduces the Niagara model, the decoder from Flash3D \cite{szymanowicz2024flash3d}, and the integration of monocular depth prediction as a prior, followed by 
a description of the 3D self-attention and geometric affine field that we use to learn better geometric features.

A detailed overview of our method is given in \autoref{fig:Overview}. Our method combines depth and normal estimation \cite{yu2022monosdf,wang2022neuris,cheng2024gaussianpro,turkulainen2024dn} with a layered Gaussian representation, allowing for robust handling of occlusions and unobserved surfaces. As we will show, the method significantly improves the geometric consistency and accuracy of single-view 3D scene reconstruction.

\subsection{Prerequisites: Monocular 3D Reconstruction} \label{sec:background}

\textbf{Scene Representation.} Similar to Flash3D~\cite{szymanowicz2024flash3d}, our method represents scenes as a collection of 3D Gaussians \cite{huang2018deepmvs,chen2025mvsplat,szymanowicz2024flash3d,hu2024metric3d,long2024wonder3d,yu2025get3dgs}. For the $i$-th Gaussian, it is characterized by a set of parameters: opacity $\sigma_i$ (range: $[0, 1)$), mean position $\mu_i \in \mathbb{R}^3$ (indicating the center of the Gaussian), covariance matrix $\Sigma_i \in \mathbb{R}^{3 \times 3}$ (describing the spread and orientation), and radiance function $c_i: \mathbb{S}^2 \rightarrow \mathbb{R}^3$ (which defines the direction-dependent color). 

For each pixel, we predict the corresponding Gaussian parameters: opacity $\sigma$, depth $d \in \mathbb{R}^+$, position offset $\Delta \in \mathbb{R}^3$,  covariance $\Sigma \in \mathbb{R}^{3 \times 3}$, a normal vector $\gamma \in \mathbb{R}^3$,  color model parameters $c \in \mathbb{R}^{3(L+1)^2}$, and $\mu_i \in\mathbb{R}^{3}$, which is the mean, where $L$ means the order of spherical harmonics, thereby providing color and lighting information for each pixel. The unnormalized Gaussian function for each component is given by:
\begin{equation}
g_i(x) = \exp\left(-\frac{1}{2} (x - \mu_i)^\top \Sigma_i^{-1} (x - \mu_i)\right).
\end{equation}
\label{eq:gaussian_func}

\noindent Spherical harmonics~\cite{ramamoorthi2001signal,sloan2023precomputed,zhou2019glosh} are employed to model colors emitted by these Gaussians,
\begin{equation}
[c_i(\nu)]_j = \sum_{l=0}^L \sum_{m=-l}^l c_{ijop} Z_{op}(\nu),
\end{equation}
where $\nu \in \mathbb{S}^2$ denotes the viewing direction, and $Z_{op}$ represents the spherical harmonics.
The opacity at the location $x \in \mathbb{R}^3$ can be derived by:
\begin{equation}
\sigma(x) = \sum_{i=1}^G \sigma_i g_i(x),
\label{eq:opacity}
\end{equation}
where $G$ represents the number of Gaussians, the radiance $c(x, \nu)$ at location $x$ from viewing direction $\nu$ is given by:
\begin{equation}
c(x, \nu) = \sum_{i=1}^G \frac{c_i(\nu) \sigma_i g_i(x)}{\sigma(x)}.
\label{eq:color}
\end{equation}
This representation captures both opacity and radiance, offering a detailed view of the scene from different angles.

\vspace{0.3em}
\noindent \textbf{Differentiable Rendering.} To render the radiance field into an image $\hat{J}$, we integrate radiances along each line of sight using the emission-absorption equation \cite{max1995optical} following 3D Gaussian Splatting~\cite{kerbl20233d}:
\begin{equation}
\begin{split}
   \hat{J} &= \text{Rend}(\mathbf{G}, \pi) \\
    &= \int_0^\infty c(x_t, \nu) \sigma(x_t) \exp\left(-\int_0^t \sigma(x_\tau) d\tau\right) dt, 
\end{split}
\end{equation}
where $\mathbf{G}$ means the Gaussian mixture, and $\pi$ means the the viewpoint; $x_t = x_0 - t\nu$ is a point from the camera center $x_0$ with distance $t$ at the direction of $\nu$.

\vspace{0.3em}
\noindent \textbf{Monocular Reconstruction.}
\label{sec:monocular}
Inspired by~\cite{szymanowicz2024splatter}, the output of the neural network $\Phi(I) \in \mathbb{R}^{C \times H \times W}$ specifies the parameters of a colored Gaussian for each pixel $u = (u_x, u_y, 1)$. The mean of each Gaussian is:
\begin{equation}
\mu = K^{-1}u d + \Delta,
\end{equation}
where $K = \text{diag}(f, f, 1) \in \mathbb{R}^{3 \times 3}$ is the camera calibration matrix with focal length $f$.

The neural network $\Phi$ is trained using image triplets consisting of $(I, J, \pi)$, where $I$ is the input image, and $J$ is the target ground-truth image at camera pose $\pi$.

\subsection{Our Method: Niagara} \label{sec:Baseline}

Our Niagara method integrates both the normal and depth as input, which are then transformed to a geometric affine field. It is then converted to 3D Gaussian parameters that can be used to render high-quality novel views.

\subsubsection{Prior Information and Geometric Feature}

\noindent \textbf{Normal-Integrated Depth Estimator.} To overcome the inaccurate geometry in previous methods (\textit{e.g.}, Flash3D, shown in \autoref{fig:Motivation}), we propose an improvement by incorporating the concept of \textit{surface normals} into our framework. The inclusion of surface normals empowers our model to improve both the photometric precision and lighting uniformity within rendered scenes. Specifically, we use predicted per-pixel normals and depth from \textit{pre-trained} normal estimator $\Phi_n$ \cite{ye2024stablenormal} and \textit{pre-trained} depth estimator $\Psi$ \cite{piccinelli2024unidepth}, yielding a normal map $N$ and depth map $D$:
\begin{equation}
\begin{split}
     &N = \Phi_n(I), \\
     &D = \Psi(I).
     \label{eq:foundation model}
\end{split}
\end{equation}
\noindent \textbf{Gaussian Splatting Geometric Feature.} The surface normals positively influence the Gaussian parameters predicted by our network. The overall output from our network can be represented as:
\begin{equation}
[\Phi(I, D, \gamma)]_u = (\sigma, \Delta, s, \theta,  c),
\label{eq:gaussian_param}
\end{equation}
where additional parameters $s$, $\theta$, $\gamma$, and $c$ mean shape, orientation, normal, and color, respectively; \( \sigma \) is the positive opacity, \( \Delta \) is the 3D displacement vector, \( s \) denotes 3D scale factors, \( \theta \) is a quaternion for rotation \( R(\theta) \), $\gamma$ is the normal vector, and \( c \) represents color parameters.
Incorporating normals improves the computation of the covariance matrix \( \Sigma \), where the \( \Sigma \) for each Gaussian is defined as
\begin{equation}
\Sigma = R(\theta)^\top \, \text{diag}(s) \, R(\theta).
\label{eq:covariance}
\end{equation}

Here, $R(\theta)$ is a rotation matrix parameterized by $\theta$, $\text{diag}(s)$ represents a diagonal matrix of scales, and $\lambda$ is a weight factor that scales the contribution of the normal vector $\gamma$. This formulation allows the model to effectively adjust the shape and spread of the Gaussian representations based on both the geometric and photometric properties of the scene, leading to more accurate rendering.

\subsubsection{Niagara Encoder}
\label{sec:Geometic}

\noindent \textbf{3D Self-Attention.} In our framework, we use self-attention \cite{vaswani2017attention,wang2018non}, which enhances geometric constraint performance by allowing geometric features directly associated with different locations to take into account spatial location information. Similar to 
 the findings in MVDream \cite{shi2023mvdream}, we observe that simple temporal attention fails to learn multi-view consistency and that content drift remains an issue even after fine-tuning on a 3D-rendered dataset. 
Therefore, we decide to use a 3D attention mechanism similar to MVDream 
(see \autoref{fig:Overview}), which produces fairly consistent images even when the view gap is huge based on our findings. Specifically, given a tensor of shape $[B, H, W, C]$, we format it to $[B, H\times W, C]$ for self-attention, where the second dimension is the sequence dimension representing the number of tokens. This way, we can also inherit all the module weights from the original 2D self-attention,
\begin{equation}
\begin{split}
     &x = \mathrm{rearrange}\bigl(x, B\,C\,H\,W \rightarrow B\,(H\,W)\,C\bigr), \\
     &x = x + \mathrm{3D SelfAttn}(x),\\
     &x = \mathrm{rearrange}\bigl(x, B\,(H\,W)\,C \rightarrow B\,C\,H\,W\bigr),
     \label{eq:3d selfattention}
\end{split}
\end{equation}
\noindent where \( x \) is the feature, \( B \), \( H \), \( W \), and \( C \) mean the batch size, height, width, and the number of channels, respectively.

\noindent \textbf{Geometric Affine Field.} In our framework, we utilize a hybrid representation that combines explicit geometry similar to TensoRF \cite{chen2022tensorf} through a point cloud $P \in \mathbb{R}^{N_P \times 3}$, which consists of $N_P$ 3D points defined by their $x$, $y$, and $z$ coordinates, and an implicit feature field encoded by a local geometric affine tensor $T \in \mathbb{R}^{3 \times C \times H \times W}$, where $C$ represents the number of feature channels and $H$ and $W$ denote the height and width of the feature maps. The local geometric affine $T$ is composed of three orthogonal geometric affines aligned with the axis: $T_{xy}$, $T_{xz}$, and $T_{yz}$, which correspond to projections onto the planes $xy$, $xz$ and $yz$, respectively. To retrieve the corresponding feature vector at any given 3D position $x$, we first project $x$ onto each of the three geometric affines to obtain the projections $p_{xy}$, $p_{xz}$, and $p_{yz}$. The final feature vector $f_t$ is then computed by performing trilinear interpolation on each of these projections and concatenating the results, which can be represented with the following equation,
{\small
\begin{equation}
\begin{split}
f_t =\; \mathrm{Aff}(T_{xy}, p_{xy})  \oplus 
\mathrm{Aff}(T_{xz}, p_{xz}) \oplus \mathrm{Aff}(T_{yz}, p_{yz}),
\end{split}
\end{equation}
}
where $\mathrm{Aff}(\cdot)$ denotes the affine interpolation operation, and $\oplus$ indicates concatenation. This scheme effectively leverages the spatial structure captured by both the point cloud and the geometric affines, allowing for enhanced feature representation in 3D space and facilitating more accurate and robust modeling of complex geometries in diverse tasks, such as depth estimation and scene reconstruction.

\subsubsection{Niagara Decoder}
\noindent \textbf{Single-View Gaussian Model.} Building upon the high-quality monocular depth predictor $\Psi$ pre-trained in Flash3D \cite{szymanowicz2024flash3d}, our framework achieves enhanced geometric consistency through iterative depth refinement modules. This depth predictor processes the input image $I$ to produce a detailed depth map $D = \Psi_d(I)$, where $D \in \mathbb{R}^+_{H \times W}$ provides depth values for each pixel in the image. 

To enhance representation in complex scenes, our model employs a \textit{depth-based Gaussian} prediction approach, predicting $K > 1$ distinct Gaussians per pixel to address the limitations of single Gaussian predictions, particularly in handling occluded regions. For an input image $I$ with its depth map $D$ and normal map $N$, the network outputs a set of parameters $\hat{P}$ for each Gaussian:
\begin{equation}
 \hat{P} = \left\{ (\sigma_i, \delta_i, \Delta_i, \Sigma_i, \gamma_i, c_i) \right\}_{i=1}^K, 
\end{equation}
where the depth of the $i$-th Gaussian is calculated as:
\begin{equation}
 d_i = d + \sum_{j=1}^i \delta_j, 
\end{equation}
where \( d = D(u) \) stands for the depth from the depth map, and $\delta_1 = 0$. The mean $\mu_i$ for the $i$-th Gaussian is given by:\begin{equation}
 \mu_i = \left( \frac{u_x d_i}{f}, \frac{u_y d_i}{f}, d_i \right) + \Delta_i,
\end{equation}
where \( f \) means the focal length of the camera. Integrating $\Delta_i$ in $\mu_i$
enables the model to better handle occluded regions and depth ambiguities.

\vspace{0.3em}
\noindent \textbf{Output Gaussian Parameters.} Using an additional neural network \( \Phi(I, D) \) to process both the image and the depth map, we output per-pixel Gaussian parameters for accurate scene modeling. For each pixel \( u \), the network result is given by Eq.~(\ref{eq:gaussian_param}), where
the mean \( \mu \) is given by
\begin{equation}
\mu = \left( \frac{u_x d}{f}, \frac{u_y d}{f}, d \right) + \Delta.
\label{eq:mean}
\end{equation}
The U-Net \cite{ronneberger2015u} with ResNet blocks \cite{he2016deep} for both encoding and decoding processes, ultimately outputs a tensor, which is $\Phi_{\text{dec}}(\Phi_{\text{enc}}(I, D)) \in \mathbb{R}^{(C-1) \times H \times W}$.
\subsubsection{Training Loss}

During the training of network $\Phi$ with data triplets $(I, J, \pi)$, the objective is to minimize the rendering loss,
\begin{equation}
    \begin{split}
\mathcal{L} = &\sum_{pm\in PM}\alpha_i ||pm(\hat{J},J)|| + 
\\ &\lambda_1 {||\Delta||^2_2} + \lambda_2 {||\text{scale}(\mathbf{G})||}.
    \end{split}
\end{equation}
Here, $\mathbf{G}$ represents the geometric feature map generated by the network from input data $I$; 
$\text{scale}(\mathbf{G})$ denotes Gaussian scales with thresholded $L_1$ regularization; $PM$ refers to photometric, a term which encompasses the measures of image quality known as PSNR, SSIM, and LPIPS which with thresholded $L_1$ regularization; $\Delta$ represents depth offsets with $L_2$
regularization, and $J$ is the target image we aim to approximate. The $L_1$ and $L_2$ loss measures are used to calculate the Euclidean distance and the sum of squares between the synthesized and target images. 
\section{Experimental Results}
\subsection{Experiment Setup}


\begin{table*}[t]
\setlength{\tabcolsep}{9pt}
\resizebox{1\linewidth}{!}{
\begin{tabular}{lccccccccc}
\toprule
\multirow{2}{*}{Method}& \multicolumn{3}{c}{5 frames} & \multicolumn{3}{c}{10 frames} & \multicolumn{3}{c}{u[-30, 30] frames} \\
\cmidrule{2-4} \cmidrule(lr){5-7} \cmidrule(lr){8-10}
         & PSNR↑   & SSIM↑   & LPIPS↓   &\hspace{6pt} PSNR↑    & SSIM↑   & LPIPS↓   &\hspace{6pt} PSNR↑      & SSIM↑      & LPIPS↓     \\ \hline
Syn-Sin \cite{wiles2020synsin}        & -       & -       & -        &\hspace{6pt} -        & -       & -        & \hspace{6pt}22.30      & 0.740      & -          \\
SV-MPI \cite{tucker2020single}         & 27.10   & 0.870   & -        &\hspace{6pt} 24.40    & 0.812   & -        & \hspace{6pt}23.52      & 0.785      & -          \\
BTS \cite{wimbauer2023behind}            & -       & -       & -        &\hspace{6pt} -        & -       & -        & \hspace{6pt}24.00      & 0.755      & 0.194      \\
Splatter Image \cite{szymanowicz2024splatter} & 28.15   & 0.894   & 0.110    & \hspace{7pt}25.34& 0.842   & 0.144    & \hspace{6pt}24.15      & 0.810      & 0.177      \\
MINE \cite{li2021mine}           & 28.45   & 0.897   & 0.111    &\hspace{6pt} 25.89    & 0.850   & 0.150    & \hspace{6pt}24.75      & 0.820      & 0.179      \\
Flash3D \cite{szymanowicz2024flash3d}        & \underline{28.46}   & \underline{0.899}   & \underline{0.100}    &\hspace{6pt} \underline{25.94}    & \underline{0.857}   & \underline{0.133}    & \hspace{6pt}\underline{24.93}& \underline{0.833}      & \underline{0.160}      \\ \midrule
Ours           &         \textbf{29.00}&         \textbf{0.904}&          \textbf{0.099}&          \hspace{7pt}\textbf{26.30}&         \textbf{0.862}&          \textbf{0.131}&            \hspace{6pt}\textbf{25.28}&            \textbf{0.836}&            \textbf{0.156}\\ \bottomrule
\end{tabular}}
  \caption{\textbf{Novel view synthesis comparison on the RealEstate10K dataset~\cite{zhou2018stereo}.} Following Flash3D~\cite{szymanowicz2024flash3d}, we evaluate our method on the in-domain novel view synthesis task. As seen, our model consistently outperforms \textit{all} existing methods across different frame counts (5 frames, 10 frames, u[-30,30] frames), in terms of PSNR, SSIM, and LPIPS. (\textbf{Best} results are in bold, \underline{second best} underlined. )
}
\vspace{-0.1cm}
\label{tab:Novel View Synthesis Performance}
\end{table*}

{\textbf{Datasets.}} We utilize the RealEstate10K (RE10K) dataset \cite{zhou2018stereo}, a comprehensive video resource intended for 3D scene reconstruction and novel view synthesis tasks. RE10K incorporates real estate videos from YouTube, featuring 67,477 scenes for training and 7,289 scenes for testing. 
%
%
Its scale and diversity facilitate the robust evaluation of model generalization performance. To ensure reliability, we randomly sample 3,205 frames from within ±30 frames. RE10K is a crucial asset in the fields of virtual reality, augmented reality, and SLAM.
Of note, we only employ \textbf{87.5\%}\footnote{Part of the original RE10K URLs expired when we downloaded the data, so we can only use the incomplete data for training in this work.} of the training data used by Flash3D, the prior SoTA method. Yet, as our results will show, the proposed method still achieves \textit{better} performance than Flash3D. 

We also compare with Flash3D under the same condition on the KITTI \cite{geiger2012we} dataset. The results are deferred to the supplementary material with detailed discussions about the comparison setups, where our method also performs better.
\vspace{-3mm}
\begin{figure}[!t]
\centering
\renewcommand{\arraystretch}{0.1} 
\begin{tabular}{c@{\hspace{0.001\linewidth}}c@{\hspace{0.001\linewidth}}c@{\hspace{0.001\linewidth}}c}
\includegraphics[width=0.239\linewidth]{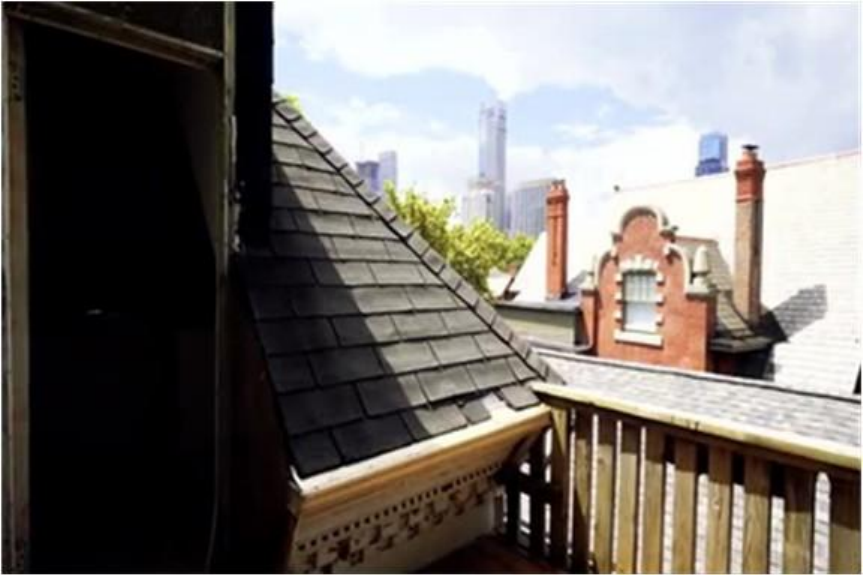} & 
\includegraphics[width=0.239\linewidth]{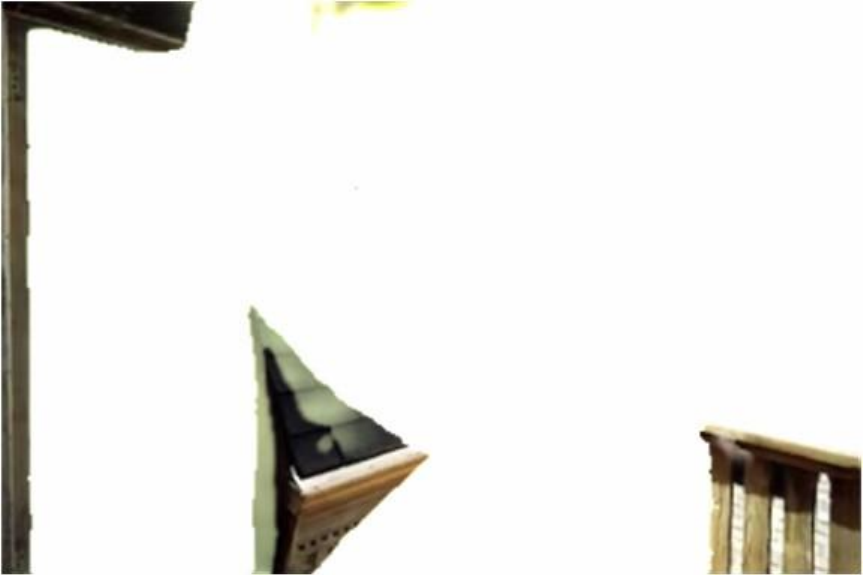} &
\includegraphics[width=0.239\linewidth]{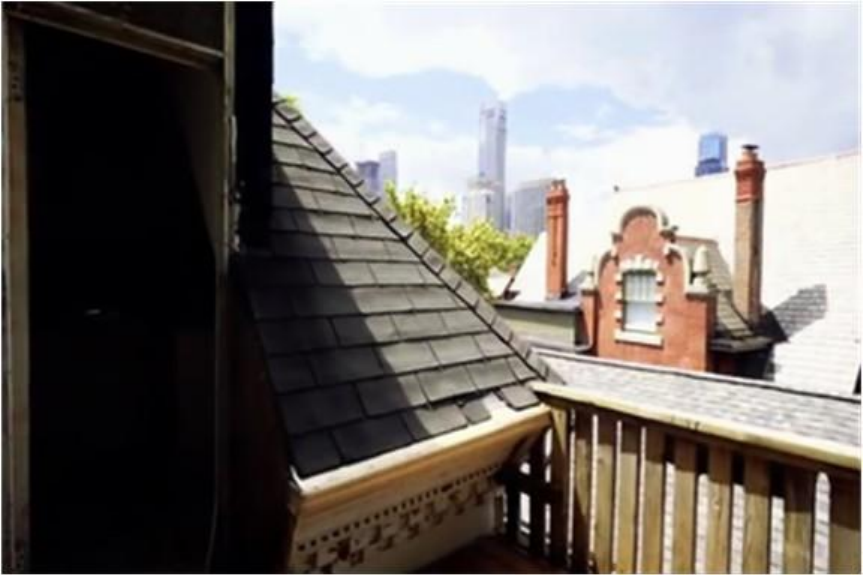} &
\includegraphics[width=0.239\linewidth]{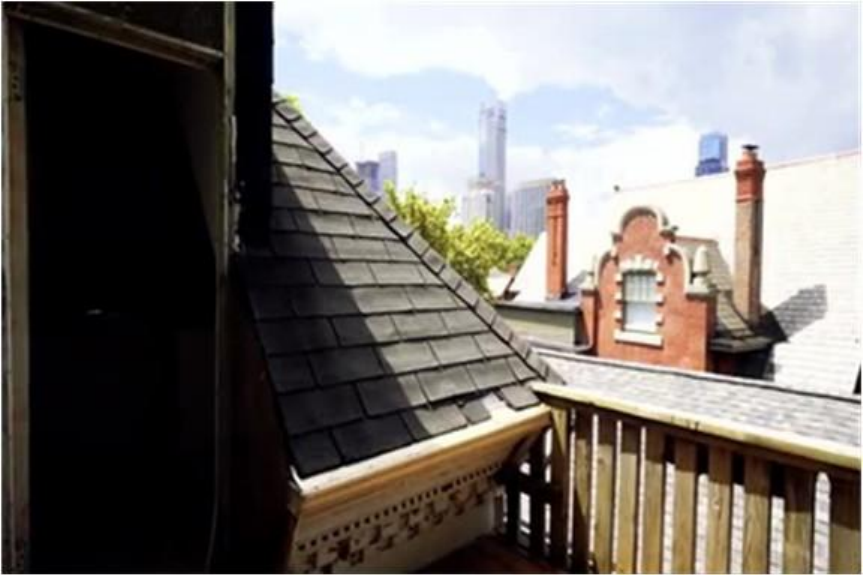} \\
\includegraphics[width=0.239\linewidth]{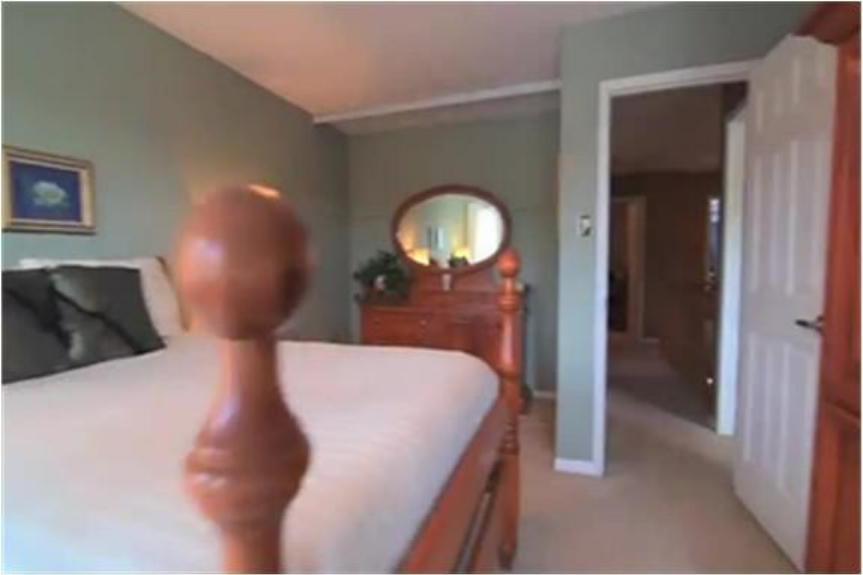} & 
\includegraphics[width=0.239\linewidth]{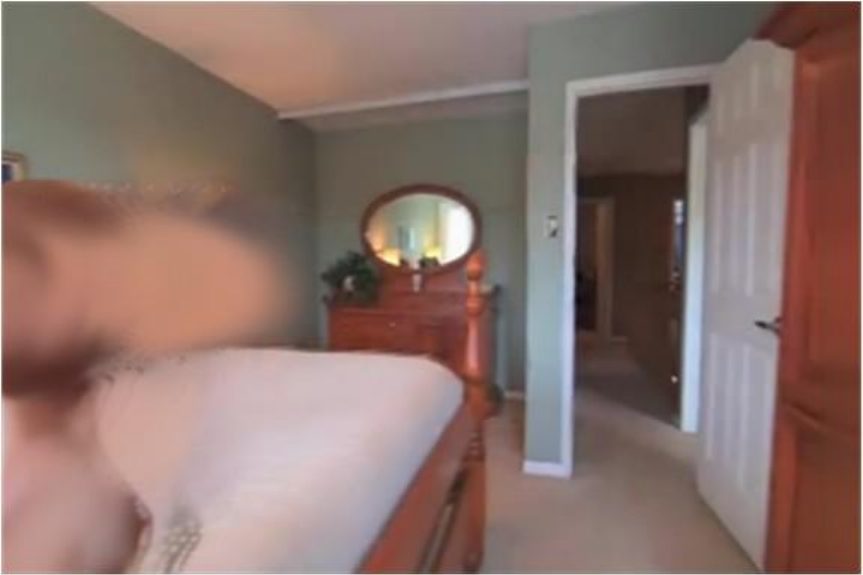} &
\includegraphics[width=0.239\linewidth]{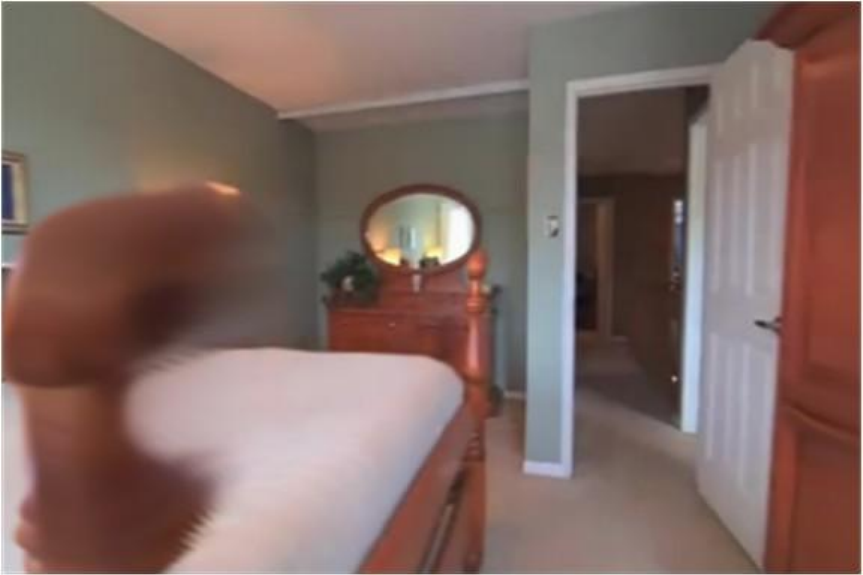} &
\includegraphics[width=0.239\linewidth]{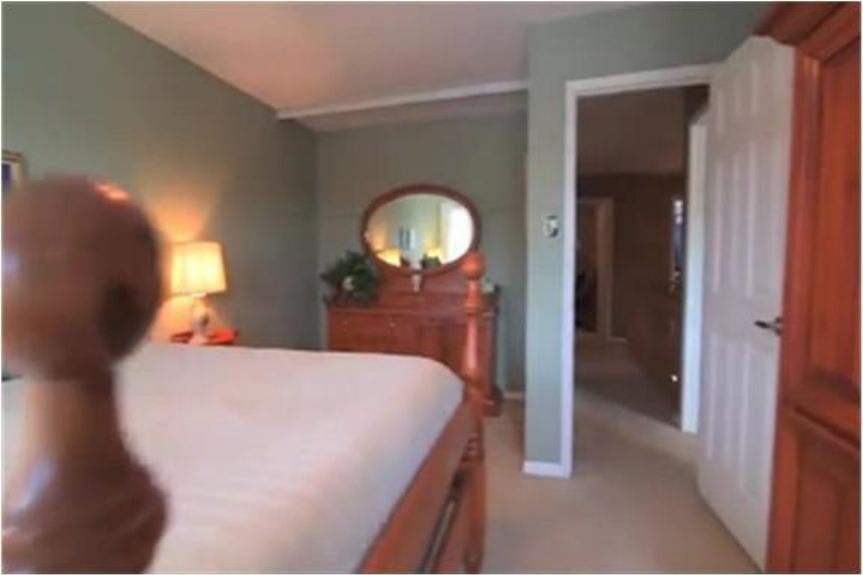} \\
\includegraphics[width=0.239\linewidth]{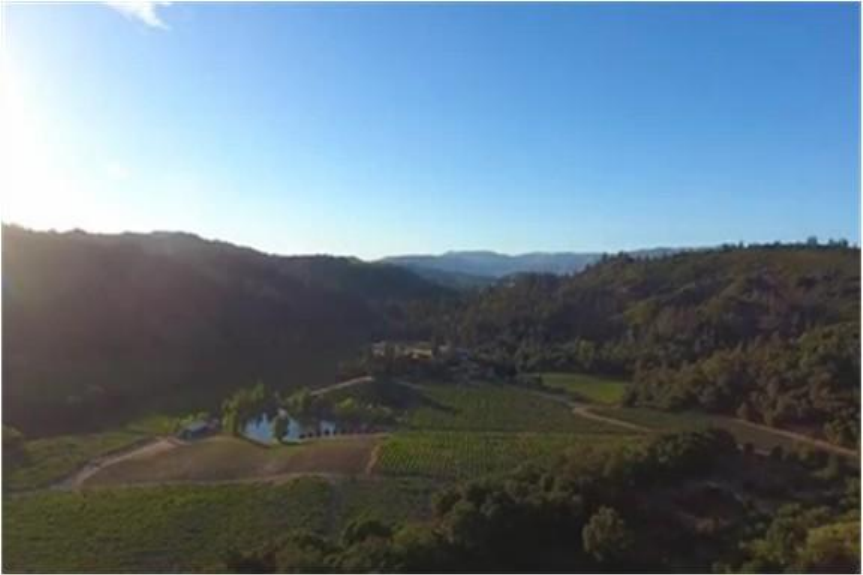} & 
\includegraphics[width=0.239\linewidth]{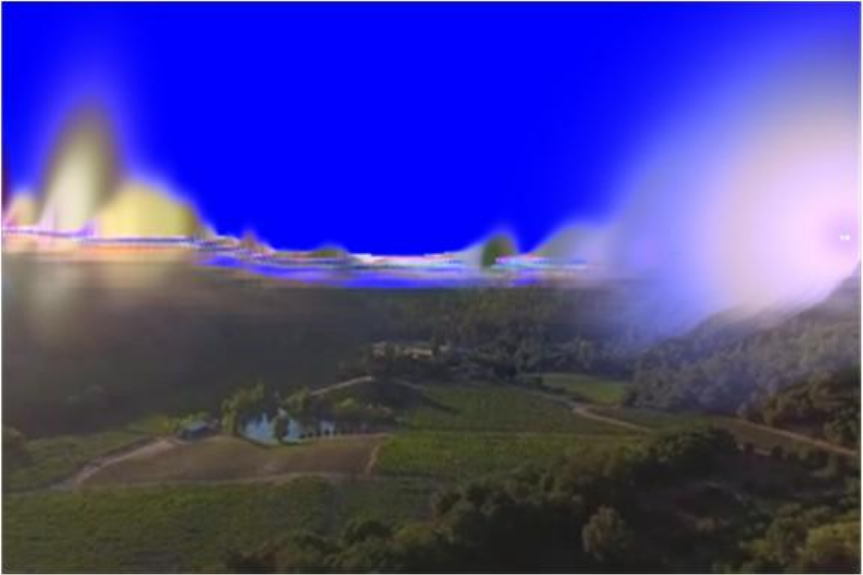} &
\includegraphics[width=0.239\linewidth]{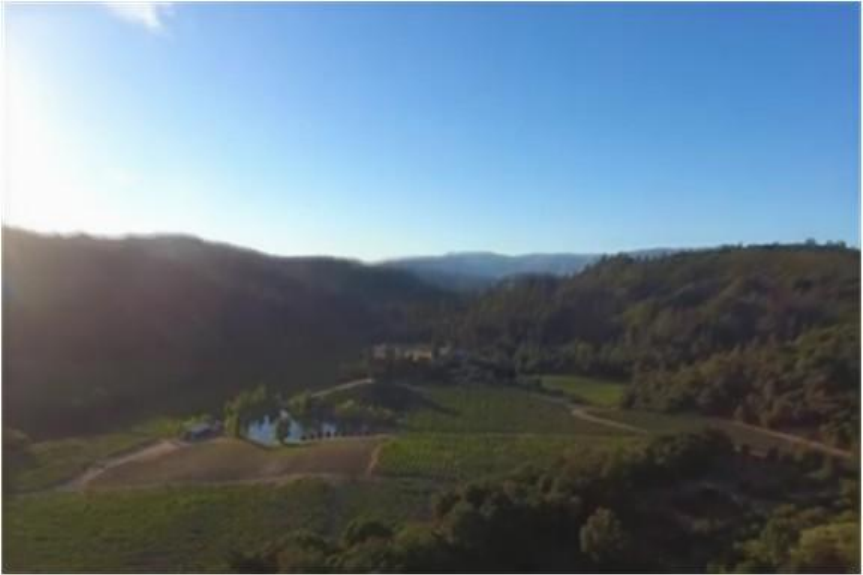} &
\includegraphics[width=0.239\linewidth]{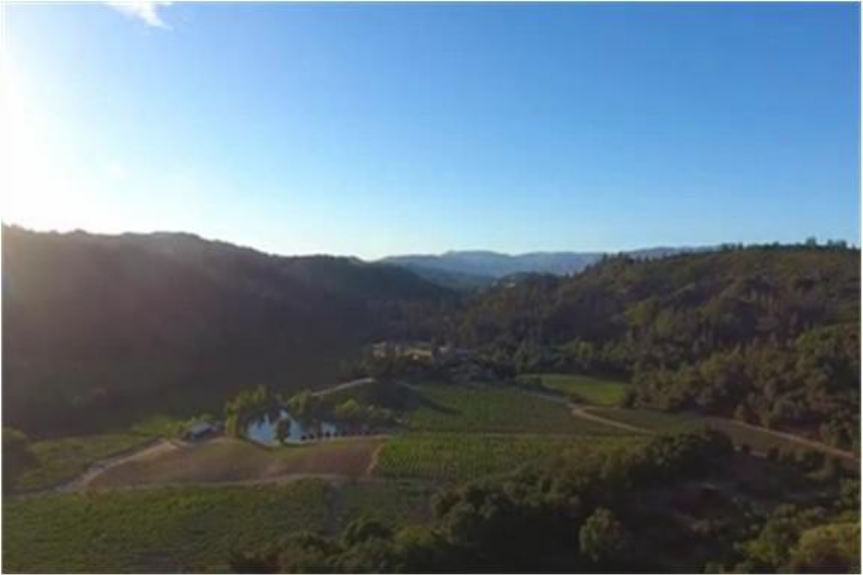} \\
\includegraphics[width=0.239\linewidth]{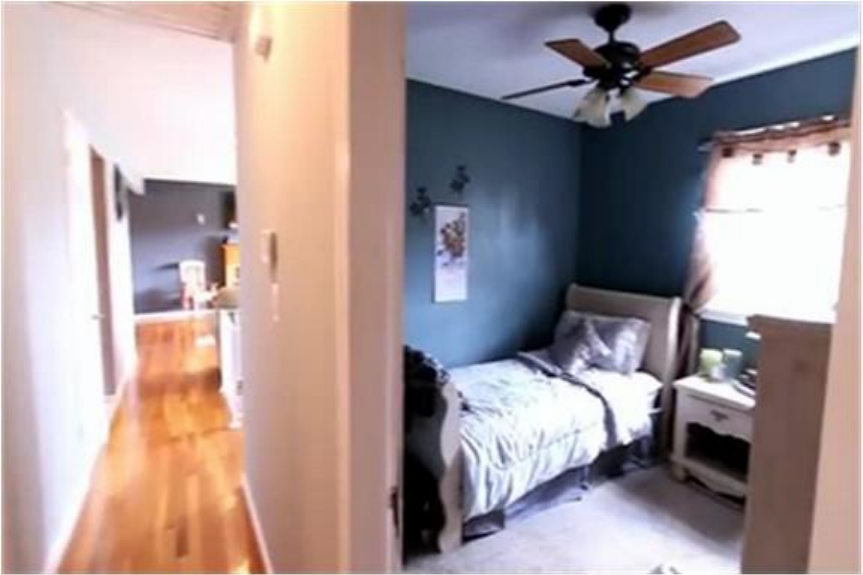} & 
\includegraphics[width=0.239\linewidth]{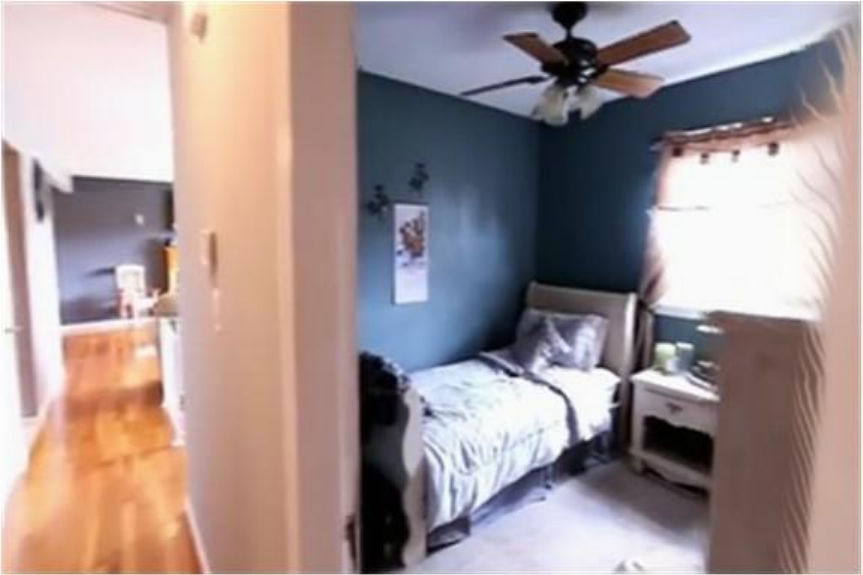} &
\includegraphics[width=0.239\linewidth]{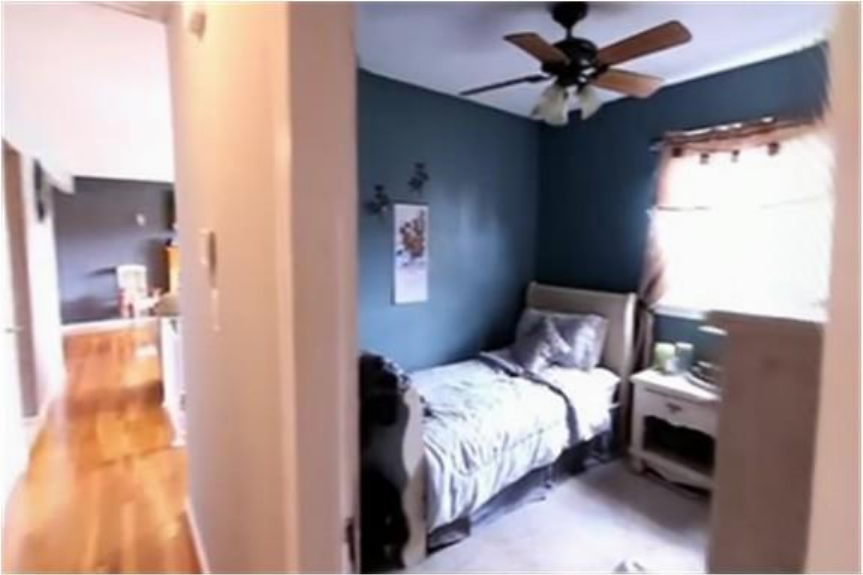} &
\includegraphics[width=0.239\linewidth]{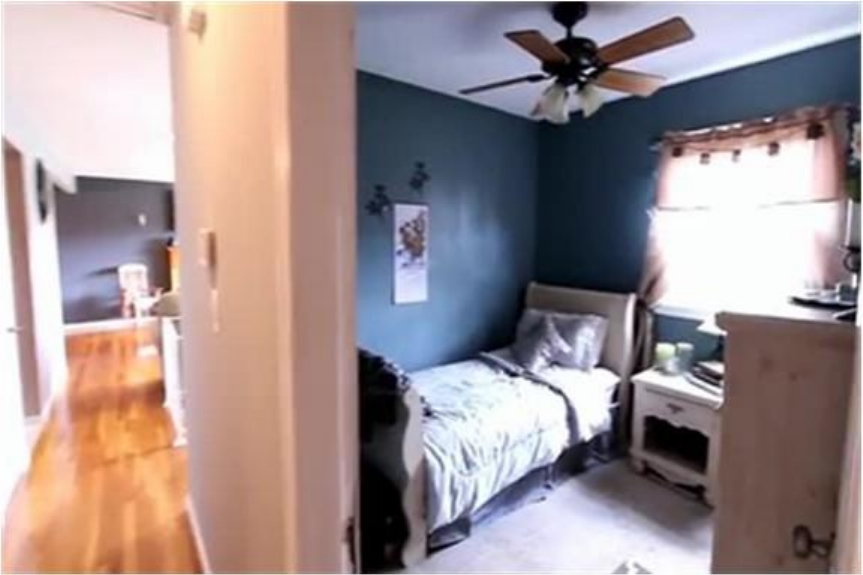} \\ 
\\ \\ 
\small (a) Input & \small (b) Flash3D & \small (c) Ours & \small (d) GT \\
\end{tabular}
\caption{\textbf{Qualitative comparison.} From top to down, the results of Row 1 are for the 5 frames setup; the results of Row 2 for the 10 frames setup; the results of Row 3 and 4 are for the u[-30, 30] frames setup. It can be observed that our Niagara mitigates the color overflow issue and produces reconstructed images with improved geometric details.}
 \label{fig:Qualitative results}
 \vspace{-0.5cm}
\end{figure}


\noindent{\textbf{Metrics and Comparison Methods.}} We evaluate the models with the MINE~\cite{li2021mine} test set partitioning and employ PSNR, SSIM, and LPIPS~\cite{castleman1996digital,wang2004image,zhang2018unreasonable} as the metrics. 
We compare our proposed method with several SoTA single-view reconstruction approaches, including  \cite{wiles2020synsin, tucker2020single,wimbauer2023behind,li2021mine,szymanowicz2024flash3d}, and an improved version of Splatter Image~\cite{westover1990footprint}. 
%
We also evaluate our method against the advanced \textit{dual-view} \cite{zitnick2000cooperative} and novel view synthesis approaches \cite{flynn2016deepstereo,sitzmann2019deepvoxels}. As we will show, despite using only a single view, our method outperforms these dual-view methods in many cases.

\vspace{2mm}
\noindent{\textbf{Training Details.}} Our Niagara employs three pretrained models: a pre-trained UniDepth model~\cite{piccinelli2024unidepth}, a pre-trained StableNormal model \cite{ye2024stablenormal}, and a ResNet50 encoder \cite{he2016deep}. Additionally, we use multiple depth offset decoders, normal offset decoders, Gaussian decoders, and geometric constraint conditions. Given the large size of the RE10K dataset, we pre-extract depth maps with UniDepth and normal maps with StableNormal. The model is trained for 50,000 iterations with a batch size of 16 on a single A6000 GPU, which takes around 26 hours.
\begin{table*}[]
\centering
\setlength{\tabcolsep}{17pt}
\resizebox{1\linewidth}{!}{
\begin{tabular}{lccccccc}
\toprule
\multirow{2}{*}{Method}& \multirow{2}{*}{Input  Views}  & \multicolumn{3}{c}{RE10K Interpolation} & \multicolumn{3}{c}{RE10K Extrapolation} \\ \cmidrule(lr){3-5} \cmidrule(lr){6-8}
& \multicolumn{1}{c}{} & PSNR↑ & SSIM↑ & LPIPS↓ & PSNR↑ & SSIM↑ & LPIPS↓ \\ \cmidrule(lr){1-2}  \cmidrule(lr){3-5} \cmidrule(lr){6-8}
Du \textit{et al.}~\cite{du2023learning} & 2 & 24.78 & 0.820 & 0.213 & 21.83 & 0.790 & 0.242 \\
pixelSplat \cite{charatan2024pixelsplat} & 2 & \underline{26.09} & \underline{0.864} & \underline{0.136} & 21.84 & 0.777 & 0.216 \\
latentSplat \cite{wewer2024latentsplat} & 2 & 23.93 & 0.812 & 0.164 & 22.62 & 0.777 & 0.196 \\
MVSplat \cite{chen2025mvsplat} & 2 & \textbf{26.39} & \textbf{0.869} & \textbf{0.128} & 23.04 & 0.813 & 0.185  \\
\hline
Flash3D \cite{szymanowicz2024flash3d} & 1 & 23.87 & 0.811 & 0.185 & \underline{24.10} & \underline{0.815} & \underline{0.185} \\
Ours & 1 & 25.24 & 0.832 & 0.162& \textbf{25.16} & \textbf{0.831} & \textbf{0.162} \\ \bottomrule
\end{tabular}}
\caption{\textbf{Comparison of different methods on the RE10K interpolation and extrapolation datasets.} For all the methods, the view closest to the target is used as the source. (1) Flash3D and ours are the only two methods that achieve \textit{single-view} scene reconstruction. Ours consistently beats Flash3D in PSNR/SSIM/LPIPS. (2) Notably, although our method utilizes only a single view, it can beat some methods using two input views, like  Du \textit{et al.}~\cite{du2023learning} and latentSplat \cite{wewer2024latentsplat}.}
\vspace{-0.5cm}
\label{tab:Interpolation and Extrapolation}

\end{table*}

\vspace{2mm}
\noindent{\textbf{Model Analyses and Ablation Studies.}}  We conduct detailed analyses of the geometric structure underlying our reconstruction method to understand its internal mechanisms. Additionally, abundant ablation studies are performed to assess the impact of each design component on the overall performance of the model, clarifying the specific contributions of each component to the model efficacy.

\subsection{Comparison Results with SoTA Methods}

\noindent {\textbf{Qualitative Comparison.}}
 \autoref{fig:Motivation} shows the results of our method with the corresponding depth maps and normal maps. The reconstructed surfaces by our method are smooth and consistent across multiviews. The appearance of the generated outdoor scenes looks natural, and indoor geometries are faithfully reconstructed. In contrast, Flash3D, due to inadequate learning of Gaussian kernel parameters, produces rainbow banding artifacts and lacks finer details in structures like tables, pillars, and walls.

\autoref{fig:Qualitative results} comparatively evaluates four reconstruction components through controlled experiments. In the 5-frame configuration (first row), Flash3D demonstrates significant rendering failures in outdoor scenes, exhibiting a complete loss of architectural geometric coherence. The 10-frame configuration (second row) reveals structural instability in Flash3D's output, manifested as anomalous edge blurring in columnar structures. Under dynamic [-30,30] frame conditions (third row), severe color bleeding artifacts occur in open environments, notably producing non-physical chromatic diffusion at vegetation-architecture boundaries. Additional tests under dynamic frame conditions (fourth row) highlight material boundary reconstruction limitations, resulting in incomplete geometric gaps at junctions between hard surfaces and transparent objects. Niagara maintains consistent structural integrity across all configurations, demonstrating enhanced surface detail sharpness while effectively eliminating color contamination.
Additional comparative analyses are provided in the supplementary material.
\begin{table*}[]
\resizebox{1.0\textwidth}{!}{
\begin{tabular}{cccccccccccccc}
\toprule
\multirow{2}{*}{Setting} & \multicolumn{4}{c}{Module} & \multicolumn{3}{c}{5 frames} & \multicolumn{3}{c}{10 frames} & \multicolumn{3}{c}{u{[}-30,30{]} frames} \\ \cmidrule(lr){2-5} \cmidrule(lr){6-8} \cmidrule(lr){9-11} \cmidrule(lr){12-14}
     & Baseline & Normal & GAF & 3D Self-Attention & PSNR ↑ & SSIM ↑ & LPIPS ↓ & PSNR ↑ & SSIM ↑ & LPIPS ↓ & PSNR ↑ & SSIM ↑ & LPIPS ↓ \\ \midrule
a    & \checkmark &        &         &                & 28.46  & 0.899  & 0.100  & 25.94  & 0.857  & 0.133  & 24.93  & 0.833  & 0.160  \\
b    & \checkmark & \checkmark      &           &       & 28.70  & 0.903  & 0.099  & 26.14  & 0.862  & 0.131  & 25.10  & 0.837  & 0.157  \\
c    & \checkmark &        & \checkmark       &      & 28.52  & 0.903  & \underline{0.098}  & 26.03  & 0.861  & 0.130  & 24.99  & 0.836  & 0.156  \\ 
d    & \checkmark & \checkmark  & \checkmark &       & \underline{28.76}  & \underline{0.904}  & \textbf{0.095} &\underline{26.16} & \underline{0.862} & \textbf{0.128} & \underline{25.16} & \textbf{0.837} & \textbf{0.154} \\ \midrule
Ours & \checkmark & \checkmark  & \checkmark & \checkmark    & \textbf{29.00}&         \textbf{0.904}&          0.099&          \textbf{26.30}&         \textbf{0.862}&          \underline{0.131}&            \textbf{25.28}&            \underline{0.836}&            \underline{0.156}\\ \bottomrule

\end{tabular}
}
 \caption{\textbf{Ablation study with quantitative results.} The three major new components introduced in our method are the normal, GAF
(Geometric Affine Field) and 3D self-attention, so here we examine their influence on quantitative performance.}
\label{tab:Ablation Study}
\end{table*}


\vspace{2mm}
\noindent {\textbf{Quantitative Comparison.}}  \autoref {tab:Novel View Synthesis Performance} presents our in-domain evaluations on RE10K. We report metrics assessing zero-shot reconstruction quality and compare our results with those from the other SoTA methods. Across different comparison setups, our method consistently achieves superior performance on \textit{all} metrics. Our method achieves higher PSNR/SSIM while significantly reducing LPIPS, ensuring both high-quality reconstruction and accurate geometry.

\begin{figure*}
\centering
\renewcommand{\arraystretch}{0.1} 
\begin{tabular}{c@{\hspace{0.001\linewidth}}c@{\hspace{0.001\linewidth}}c@{\hspace{0.001\linewidth}}c@{\hspace{0.001\linewidth}}c}

\includegraphics[width=0.194\linewidth]{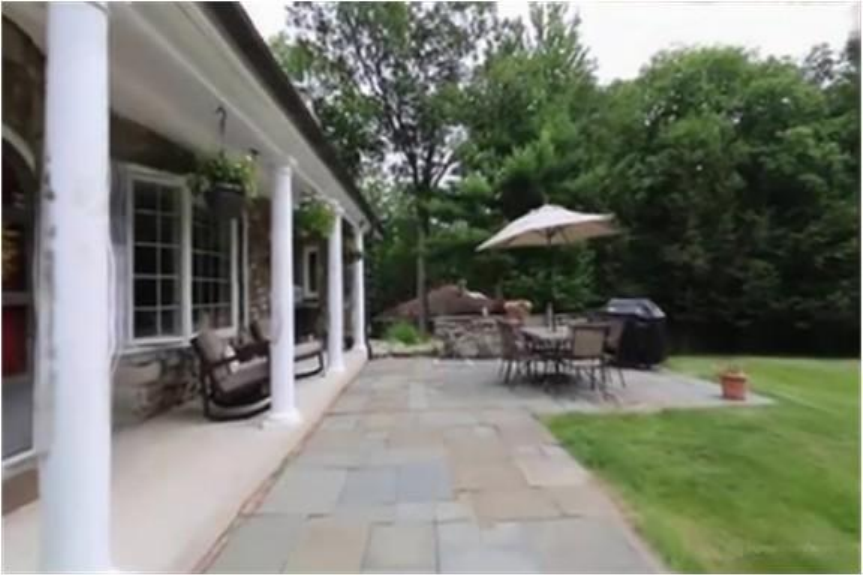} & 
\includegraphics[width=0.194\linewidth]{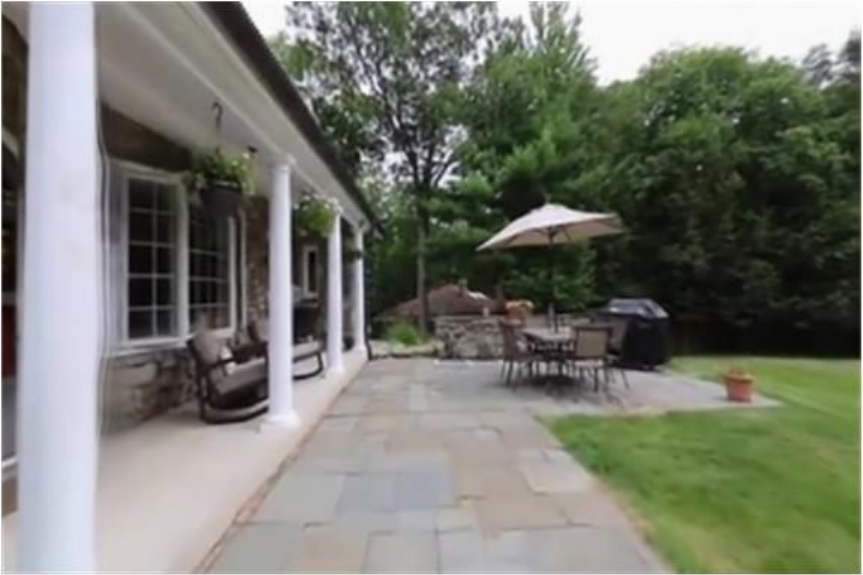} &
\includegraphics[width=0.194\linewidth]{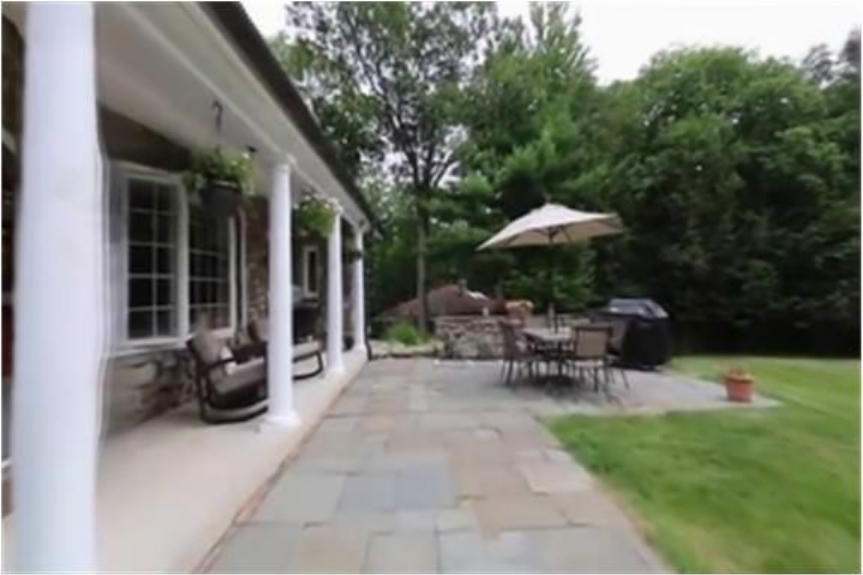} &
\includegraphics[width=0.194\linewidth]{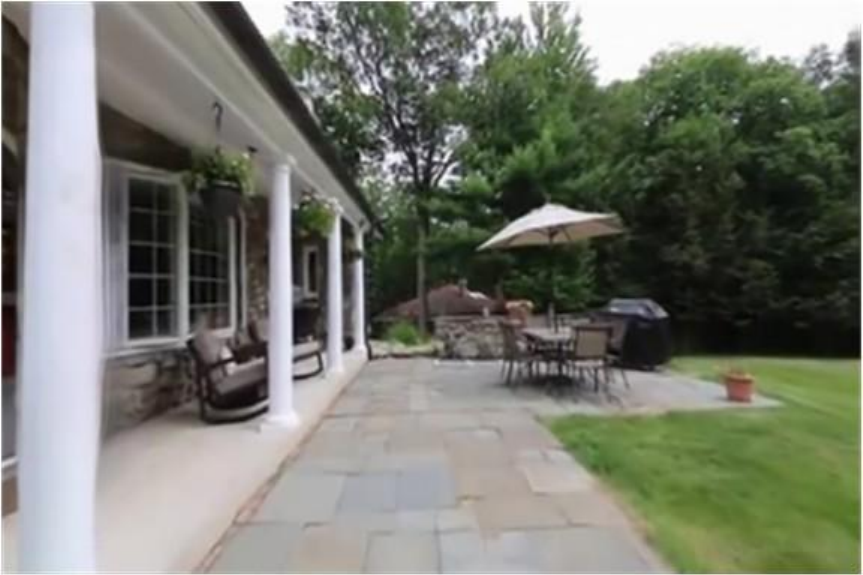} &
\includegraphics[width=0.194\linewidth]{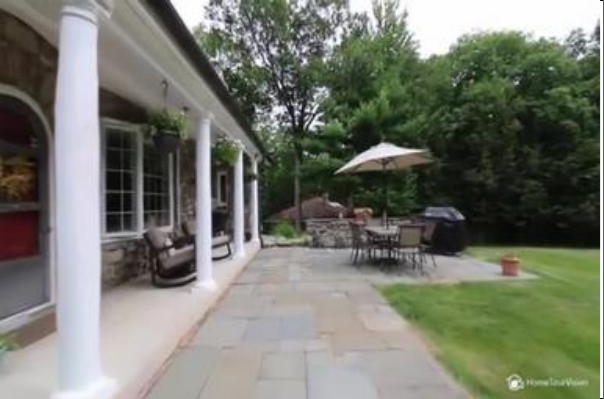} \\
\includegraphics[width=0.194\linewidth]{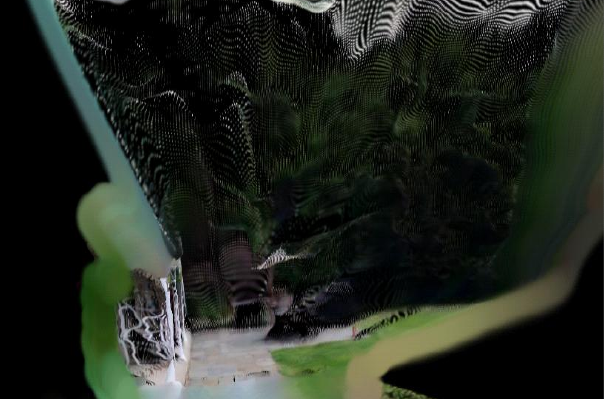} & 
\includegraphics[width=0.194\linewidth]{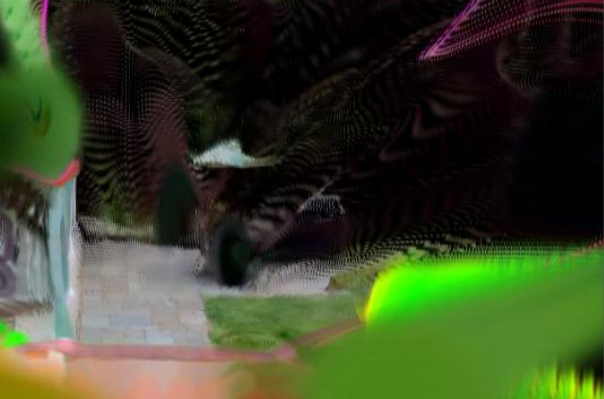} &
\includegraphics[width=0.194\linewidth]{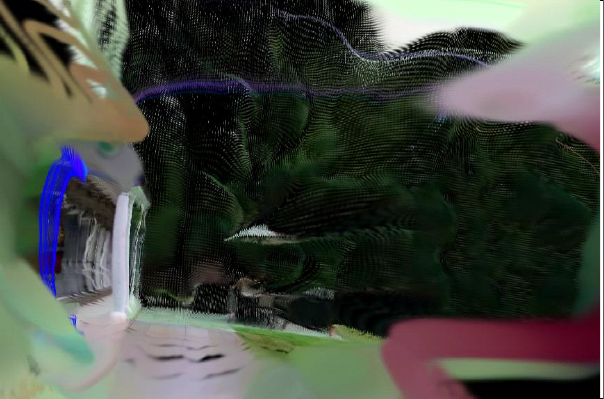} &
\includegraphics[width=0.194\linewidth]{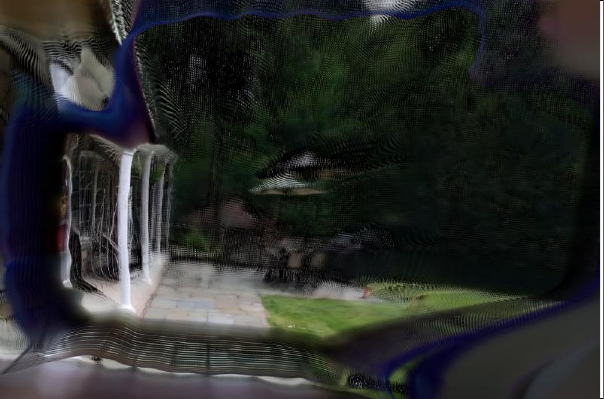} &
\includegraphics[width=0.194\linewidth]{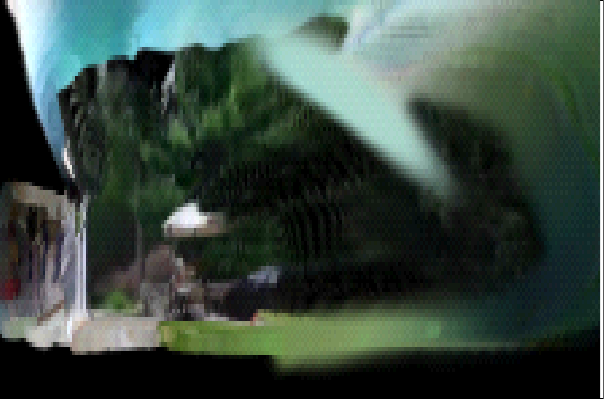} \\
\includegraphics[width=0.194\linewidth]{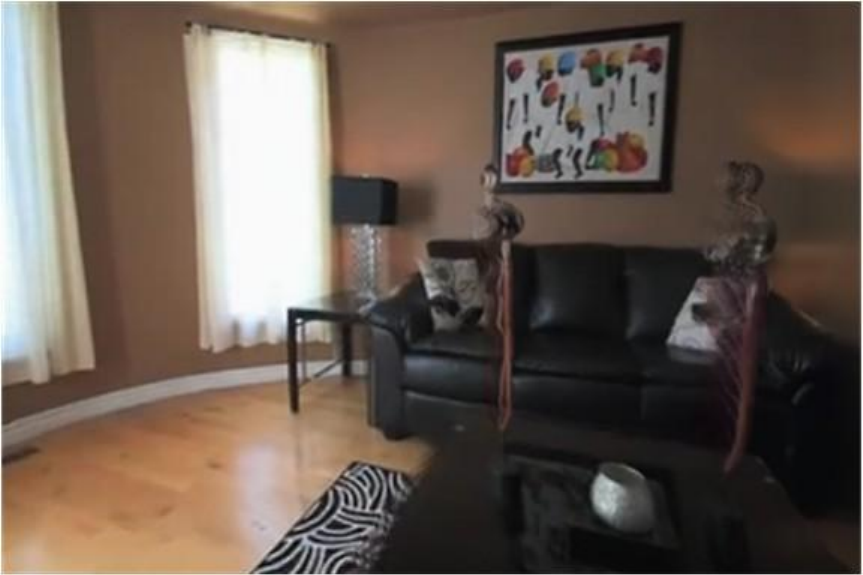} & 
\includegraphics[width=0.194\linewidth]{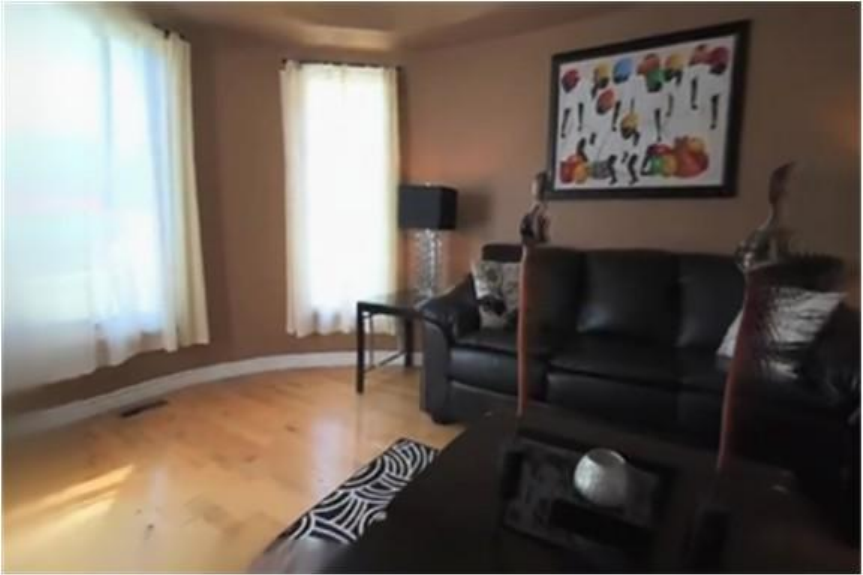} &
\includegraphics[width=0.194\linewidth]{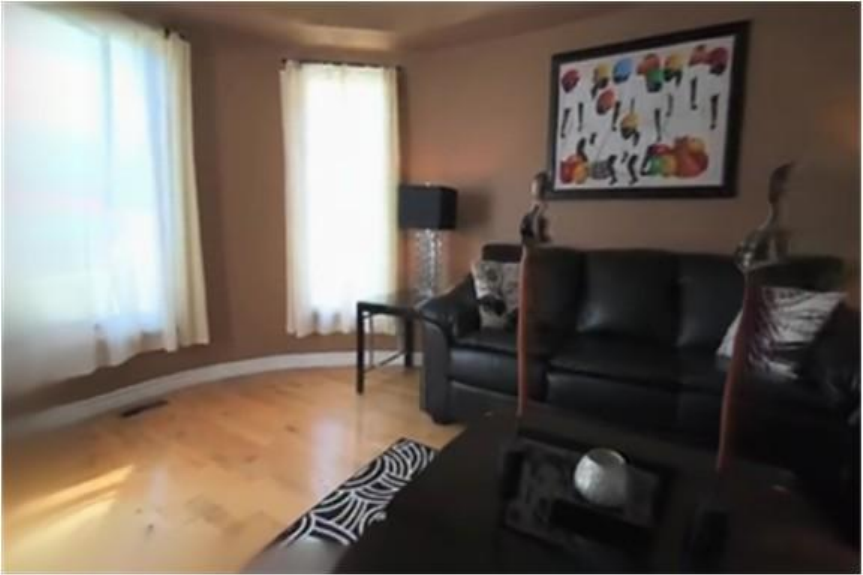} &
\includegraphics[width=0.194\linewidth]{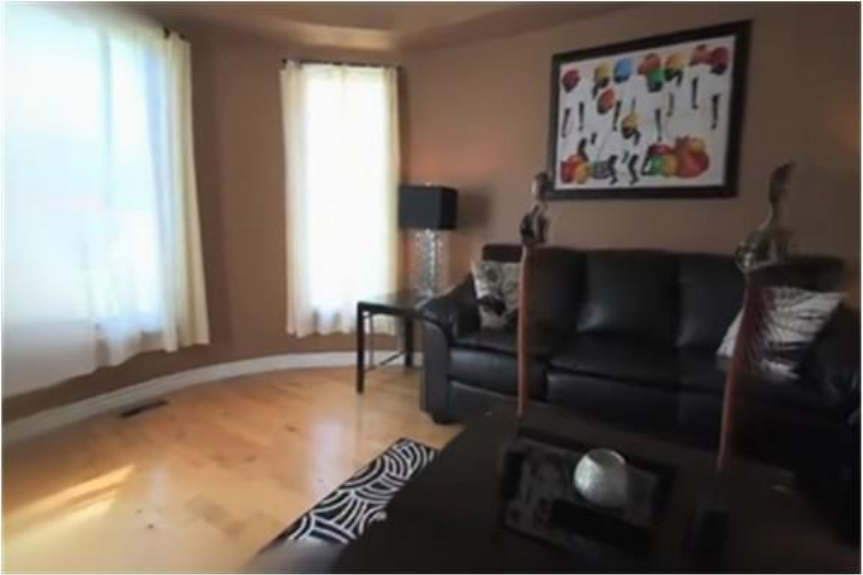} &
\includegraphics[width=0.194\linewidth]{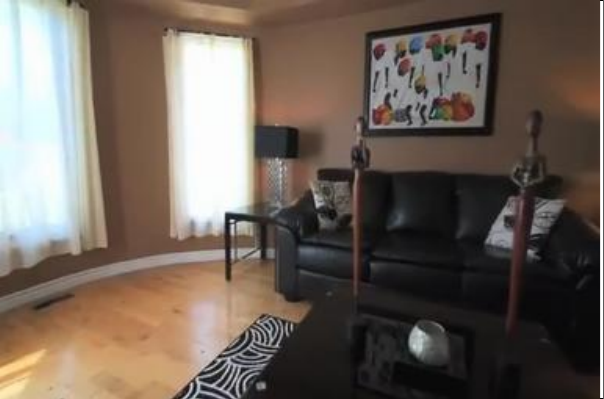}  \\
\includegraphics[width=0.194\linewidth]{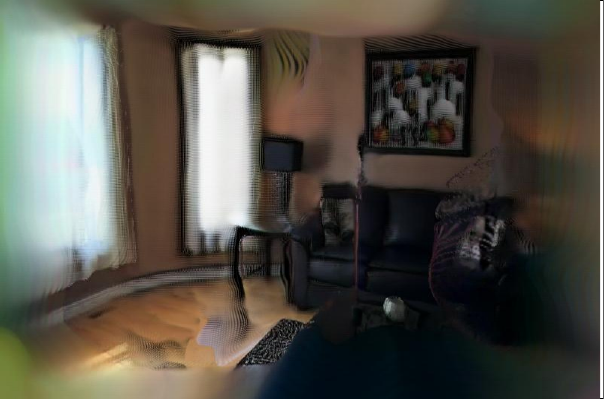} & 
\includegraphics[width=0.194\linewidth]{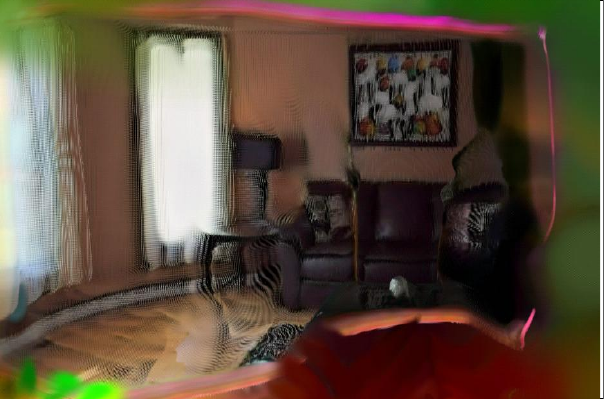} &
\includegraphics[width=0.194\linewidth]{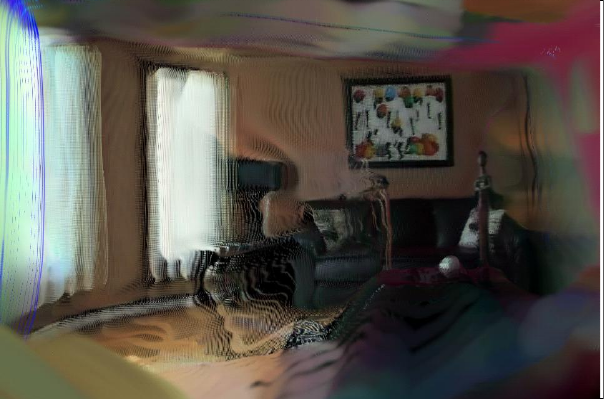} &
\includegraphics[width=0.194\linewidth]{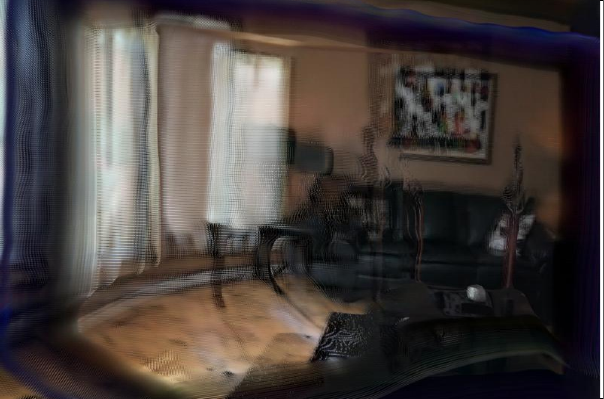} &
\includegraphics[width=0.194\linewidth]{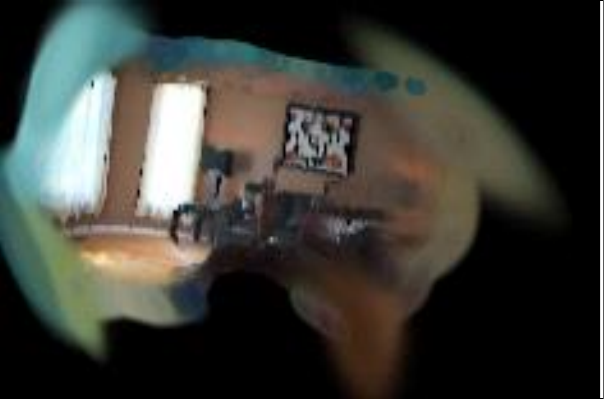}  \\
\\ \\
\small (a) Flash3D~\cite{szymanowicz2024flash3d} & \small (b) w/o Normal  & \small (c) w/o GAF  & \small (d) w/ GAF, Normal  & \small (e) Ours
\end{tabular}
\caption{\textbf{Ablation study with visual results.} All models are built upon the “base” model, which only utilizes the depth information. GAF means Geometric Affine Field, and ours adds 3D self-attention based on (d).}
\label{fig:Ablation views}
\vspace{-0.2cm}
\end{figure*}

To further evaluate the effectiveness of Niagara, we perform interpolation using pixelSplat \cite{charatan2024pixelsplat} and extrapolation using latentSplat \cite{wewer2024latentsplat}. For fairness, unlike the existing two-view methods that typically assess interpolation between two source views, Flash3D \cite{szymanowicz2024flash3d} consistently performs extrapolation from a single view. \autoref{tab:Interpolation and Extrapolation} shows the results of comparing our method with these two-view approaches and Flash3D. Although Niagara does not surpass two-view methods in interpolation due to the input disadvantage of using only one view 
it still achieves a notable \textbf{10.8\%}  LPIPS reduction over Flash3D. Additionally, Niagara excels in the extrapolation task, significantly outperforming \textit{all} prior two-view methods and demonstrating substantial improvement over Flash3D, with a \textbf{10.8\%}  LPIPS reduction.


\vspace{-0.1cm}
\subsection{\textbf{Ablation Study }}
\textbf{\textbf{Quantitative ablation study.}} In this ablation study, we evaluate the impact of critical new components in our method by comparing performance metrics presented in  \autoref{tab:Ablation Study}. The results demonstrate significant improvements across all metrics with the addition of normal and geometric constraints when contrasted with Flash3D \cite{szymanowicz2024flash3d} and highlight the essential role of these components in enhancing model performance. 

Specifically, the incorporation of normal constraints significantly enhances the overall quality of scene reconstructions by enabling the model to better capture and express subtle surface variations with improved fidelity. In addition, geometric constraints encourage 
the accurate representation of fine geometrical details, ensuring that the model faithfully reproduces the intricate shapes and structures of objects. Together, these constraints work synergistically with depth information, substantially increasing the model’s capacity to learn and represent complex details and geometric structures. However, excessive geometric constraints impair the 3D self-attention module's perceptual-imaginative capabilities, degrading LPIPS/SSIM. 

\vspace{1mm}
\noindent \textbf{\textbf{Qualitative ablation study.}}  We further conduct ablation studies with qualitative results. In \autoref{fig:Ablation views}, the 1st and 3rd rows display rendered images, while the 2nd and 4th rows present their corresponding Gaussian Splatting representations. The top two rows depict an outdoor scene, and the bottom two rows feature an indoor scene. Our ablation study shows that reconstruction quality progressively improves with the incremental integration of the normal, the geometric affine field, and the 3D self-attention module. \textbf{(1)} On the outdoor scene, the sequential incorporation of these modules yields sharper outputs, enhanced color fidelity, and improved feature alignment. \textbf{(2)} On the indoor scene (see the bottom two rows), the inclusion of these components results in more accurate geometric details, finer textures, and better-aligned features. These findings indicate that the synergistic combination of the normal module, geometric affine field, and 3D self-attention substantially elevates reconstruction fidelity. Specifically, our model achieves superior geometric precision and texture reproduction, thus enabling faithful reconstructions of complex real-world scenes. 

 \section{Conclusion}

This paper introduces \textit{Niagara}, the \textit{first} comprehensive single-view 3D reconstruction framework tailored for complex \textit{outdoor} scenes. At its core, Niagara integrates surface normals into a depth-based reconstruction pipeline to capture finer geometric details, while incorporating a geometric affine field and a 3D self-attention module for robust spatial constraint enforcement. Empirically, on the RE10K benchmark, Niagara demonstrates encouraging performance, outperforming the prior SoTA approach Flash3D~\cite{szymanowicz2024flash3d} with consistent PSNR, SSIM, and LPIPS improvements.
In the RE10K interpolation and extrapolation settings, our single-view method surpasses Flash3D by over 1 dB in PSNR and even outperforms those methods using two views under the interpolation setup.
This work establishes practical foundations for advancing \textit{single-view} 3D reconstruction for geometrically complex outdoor scenes.


\section*{Acknowledgement}
This paper is supported by Science and Technology Project of Zhejiang Province (No. 2025C01026) and Scientific Research Project of Westlake University (No. WU2025WF003).

\appendix

\section{Implementation Details}  
In this section, we describe the implementation details of our method, including the network architecture and training hyperparameters.  

\noindent\textbf{Backbone Structure.}  
We utilize a ResNet-based architecture inspired by UniDepth2 \cite{piccinelli2024unidepth}, specifically employing ResNet50 as the backbone. Custom modifications include 64 channels for feature extraction and maintaining resolution during the upsampling stage, drawing inspiration from the Flash3D \cite{szymanowicz2024flash3d} architecture. Pre-trained weights are used to enhance initial performance.  

\noindent\textbf{Geometric Affine Field.}  
The geometric affine field has a plane size of 32 with 64 channels, enabling robust spatial feature representation for complex scene rendering.  

\noindent\textbf{UniDepth Model.}  
The UniDepth model incorporates the Vision Transformer (ViT-L/14), which is known for its strong contextual representation capabilities.  

\noindent\textbf{Depth Model.}  
The depth estimation model adopts a ResNet-style architecture with 50 layers. Decoder layers are designed to process feature maps with progressively increasing channels (32, 32, 64, 128, 256), ensuring the capture of both fine and coarse details. The design includes pre-batch normalization and random background color augmentation within the Gaussian rendering pipeline, improving robustness across diverse scenarios.  

\noindent\textbf{3D Self-Attention.}
The 3D self-attention module employs 8 attention heads with 64-dimensional hidden space per head (attn\_heads: 8, attn\_dim\_head: 64), stacked across 2 transformer layers (attn\_layers: 2). This design captures long-range spatiotemporal dependencies while maintaining computational efficiency.

\noindent\textbf{Multi-frame and Gaussian Handling.}  
The model processes frames (-1, 0, 1, 2) to estimate depth, utilizing two Gaussians per pixel to refine density and depth predictions. Gaussian rendering is employed to achieve high-precision visual outputs. Additionally, pose information is integrated with rendering, further improving performance.  

\noindent\textbf{Hyperparameters.}  
The training pipeline for the RE10K dataset is optimized using a consistent set of hyperparameters across multiple GPUs. A batch size of 16 with 16 data loader workers ensures efficient input throughput. The learning rate is set to 0.0001, with mixed precision (16-bit) training to optimize computational resources. Training runs for one epoch with a step scheduler reducing the learning rate every 5,000 steps. Models are checkpointed every 5,000 iterations, with progress logged every 250 steps. To balance storage and tracking, a maximum of five checkpoints is retained. Exponential Moving Average (EMA) updates occur every 10 steps after an initial buffer of 100 steps. Depth accuracy is improved by scaling poses based on estimated depths.  

\noindent\textbf{Loss Configuration.}  
The loss function integrates multiple components to achieve a balance between spatial precision and perceptual quality. Gaussian scales and offsets are weighted to regulate spatial representation accuracy and ensure fine detail preservation. A combination of PSNR, SSIM, and LPIPS ensures pixel-wise accuracy and perceptual consistency, with SSIM and LPIPS activated after sufficient training steps to refine both visual fidelity and structural coherence.  

\noindent\textbf{Dataset Specification.}  
The RE10K dataset is used for training, adhering to its original split. Data preprocessing includes normalization, depth handling, and normal map processing. Additional features streamline dataset preparation, while comprehensive dilation and augmentation strategies address dataset variability, ensuring robust training.

\section{More Qualitative Comparison}
\begin{figure*}
\centering
\renewcommand{\arraystretch}{0.1} 
\begin{tabular}{c@{\hspace{0.0015\linewidth}}c@{\hspace{0.0015\linewidth}}c@{\hspace{0.0015\linewidth}}c@{\hspace{0.0015\linewidth}}c@{\hspace{0.0015\linewidth}}c}
\multicolumn{6}{c}{\textbf{Indoor:}} \\[2mm]
\includegraphics[width=0.199\linewidth]{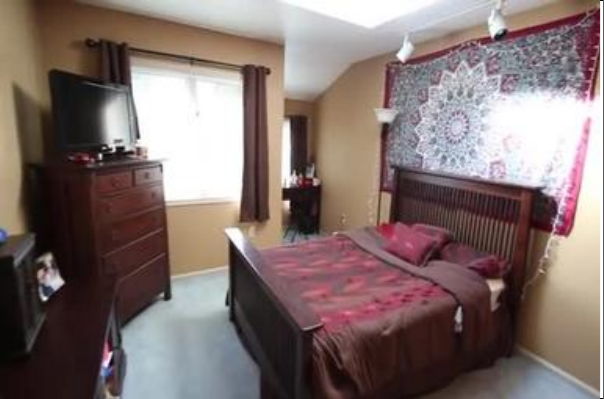} & 
\includegraphics[width=0.199\linewidth]{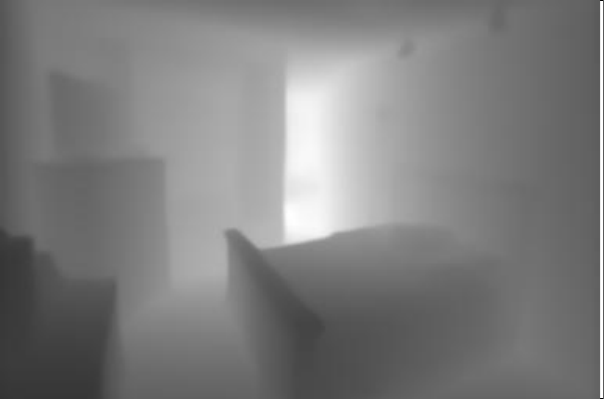} &
\includegraphics[width=0.199\linewidth]{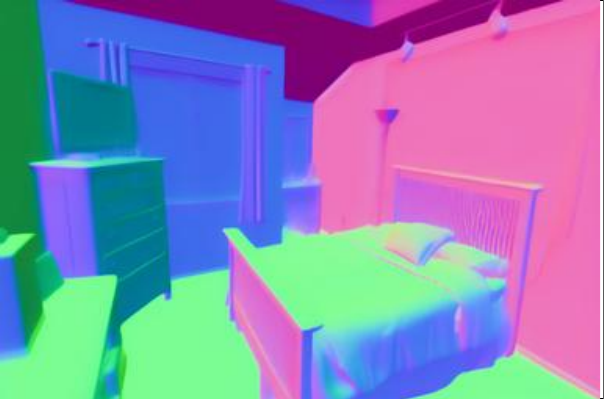} &
\includegraphics[width=0.199\linewidth]{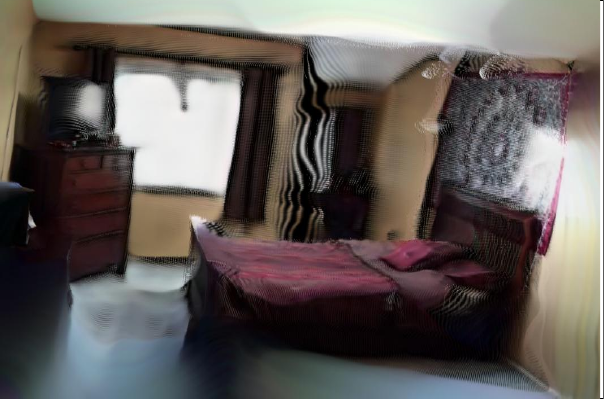} &
\includegraphics[width=0.199\linewidth]{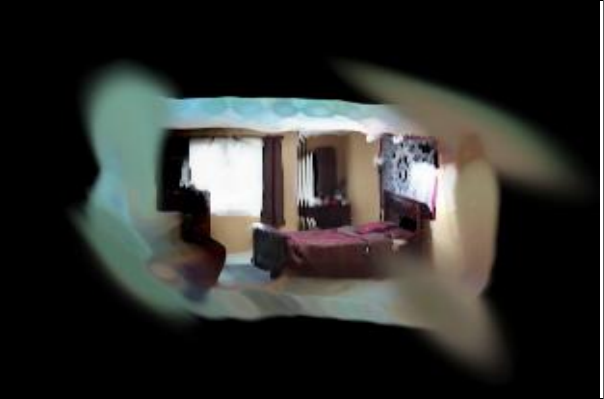} & \\
\includegraphics[width=0.199\linewidth]{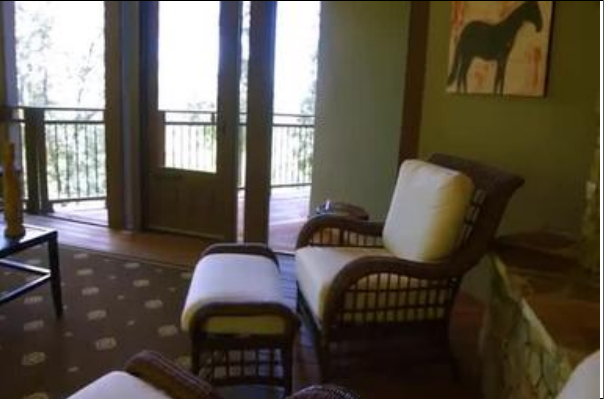} & 
\includegraphics[width=0.199\linewidth]{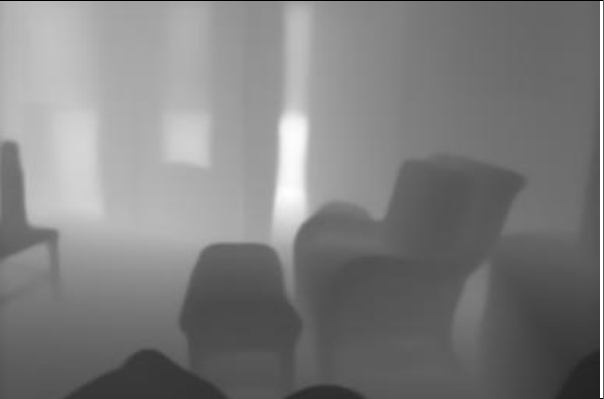} &
\includegraphics[width=0.199\linewidth]{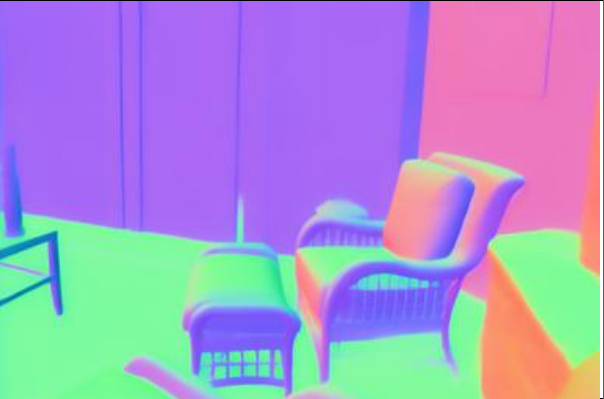} &
\includegraphics[width=0.199\linewidth]{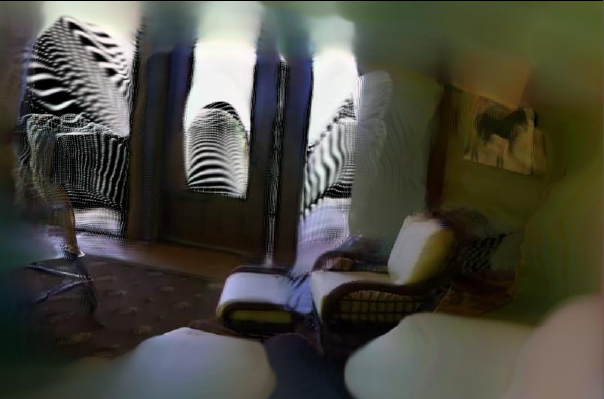} &
\includegraphics[width=0.199\linewidth]{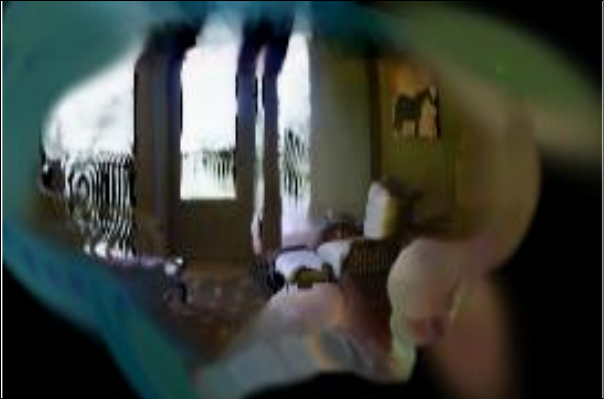} & \\
\includegraphics[width=0.199\linewidth]{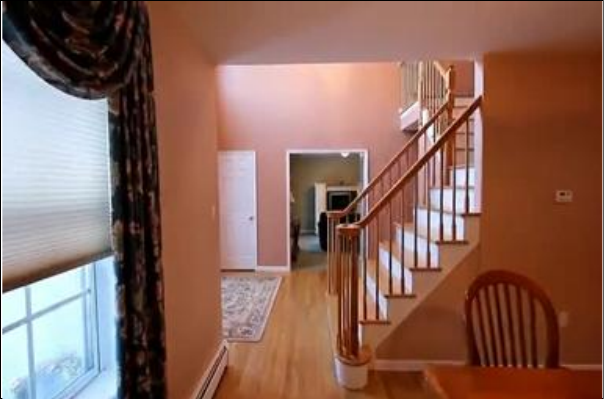} & 
\includegraphics[width=0.199\linewidth]{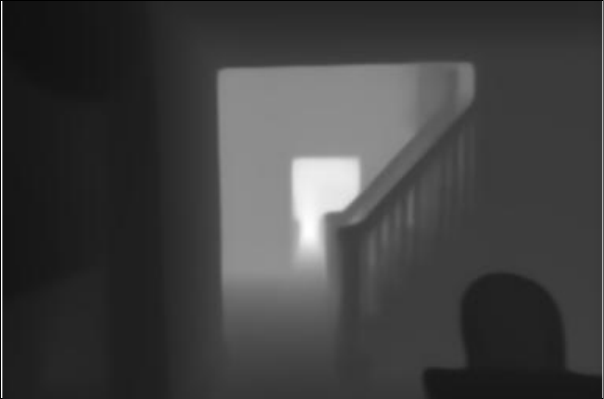} &
\includegraphics[width=0.199\linewidth]{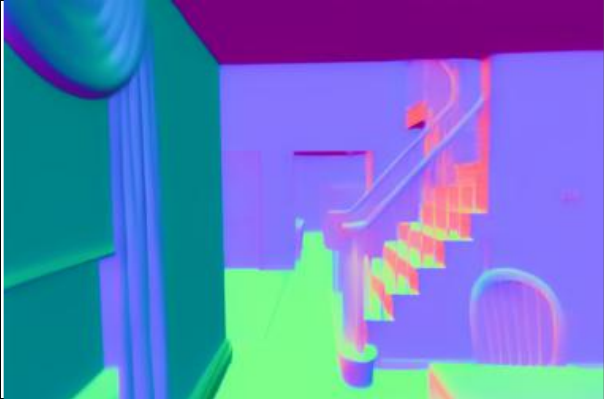} &
\includegraphics[width=0.199\linewidth]{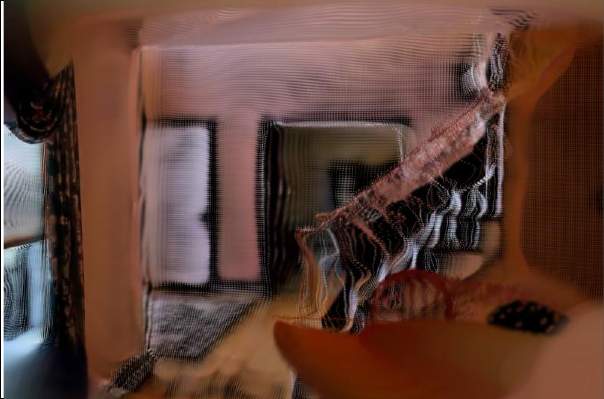} &
\includegraphics[width=0.199\linewidth]{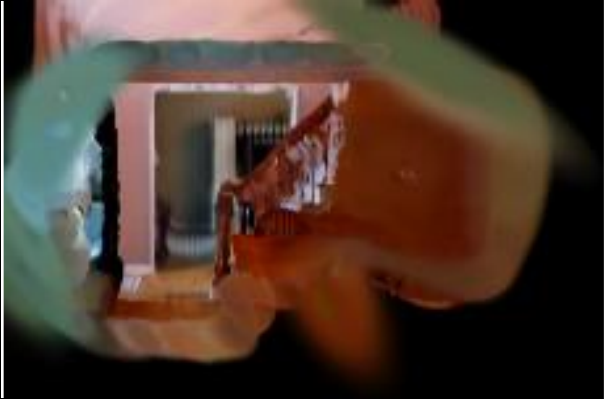} & \\
\includegraphics[width=0.199\linewidth]{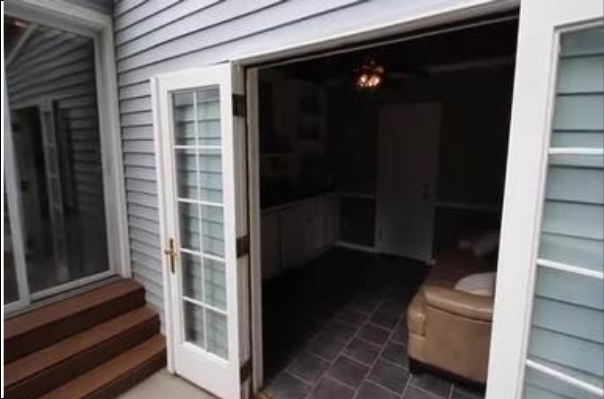} & 
\includegraphics[width=0.199\linewidth]{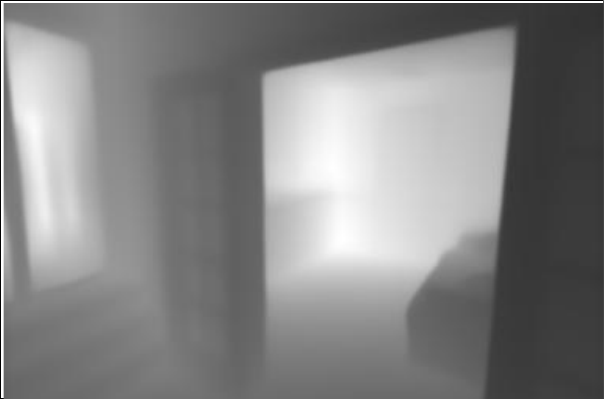} &
\includegraphics[width=0.199\linewidth]{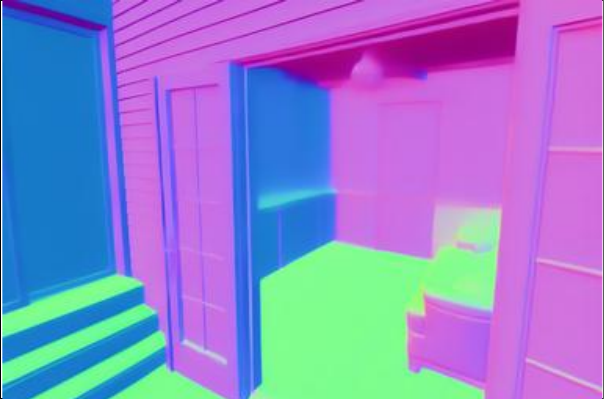} &
\includegraphics[width=0.199\linewidth]{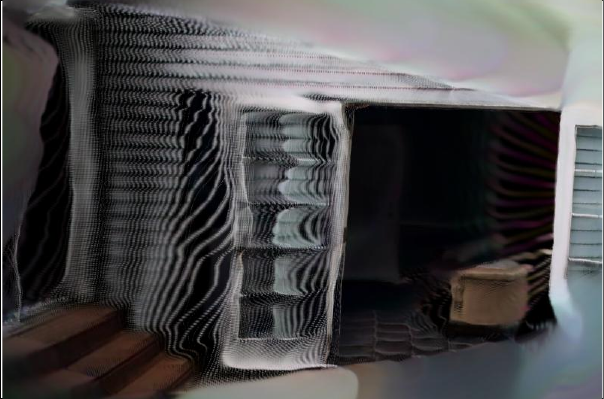} &
\includegraphics[width=0.199\linewidth]{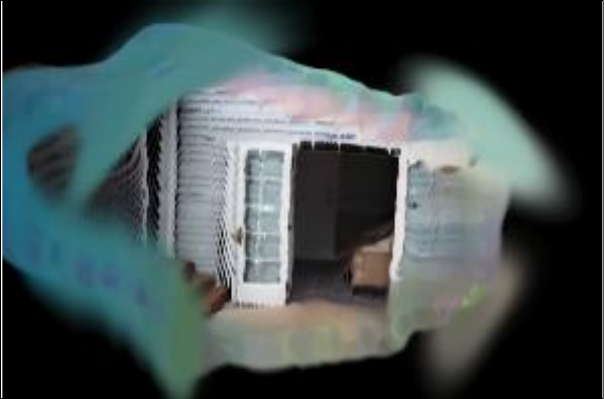} & \\
\includegraphics[width=0.199\linewidth]{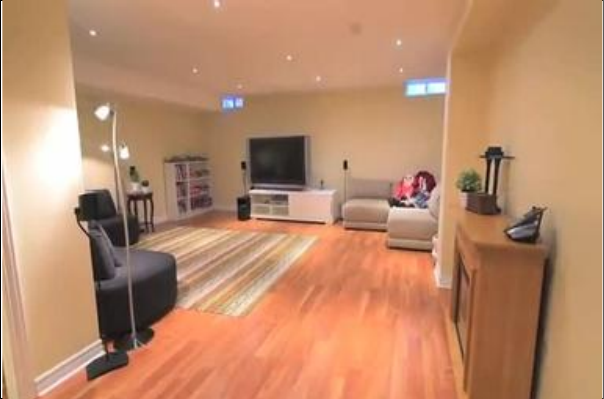} & 
\includegraphics[width=0.199\linewidth]{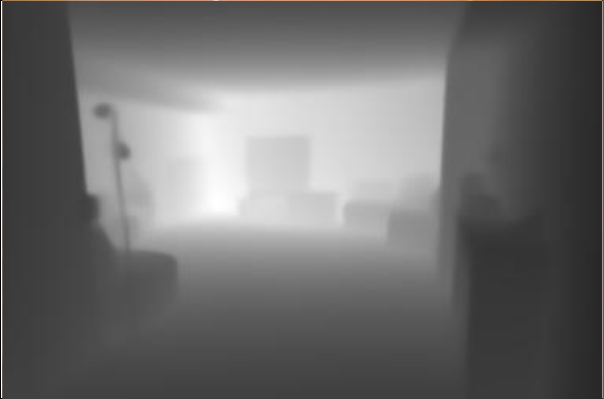} &
\includegraphics[width=0.199\linewidth]{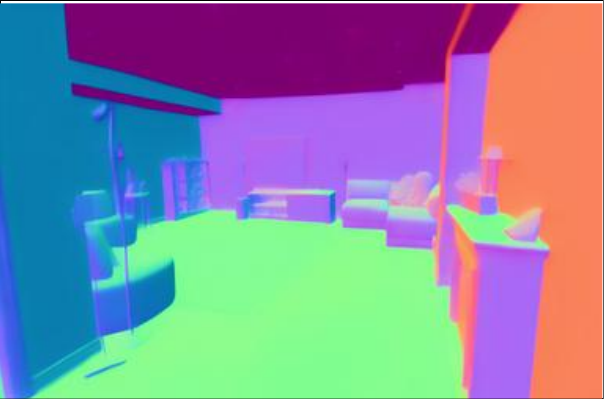} &
\includegraphics[width=0.199\linewidth]{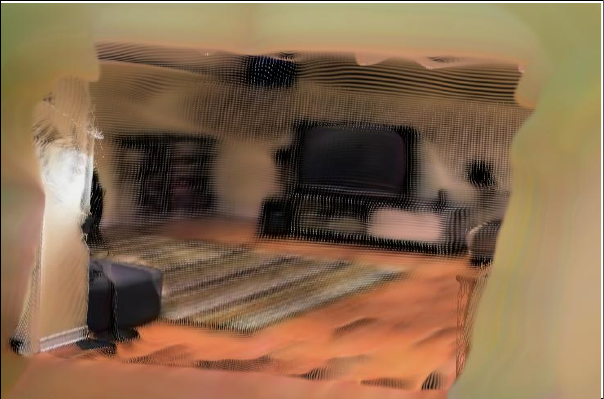} &
\includegraphics[width=0.199\linewidth]{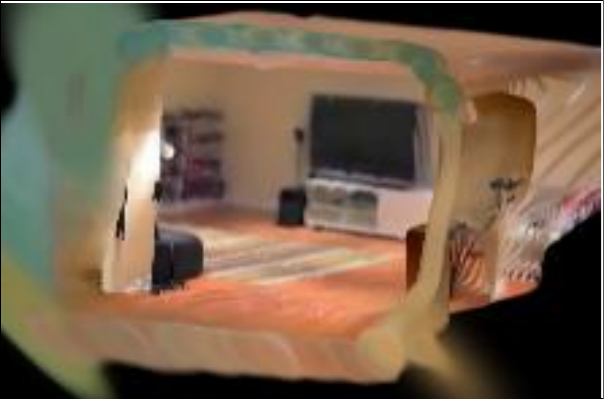} & \\
\multicolumn{6}{c}{\textbf{Outdoor:}} \\[2mm] 
\includegraphics[width=0.199\linewidth]{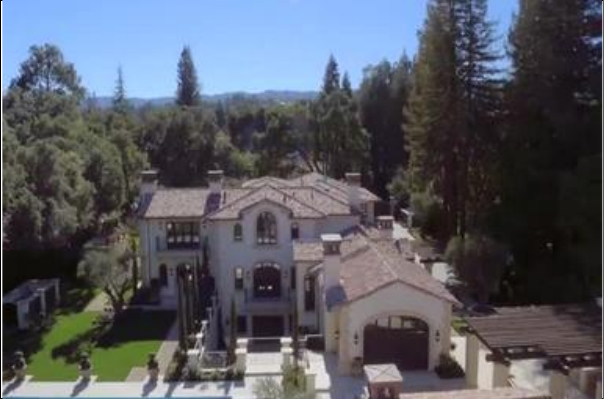} & 
\includegraphics[width=0.199\linewidth]{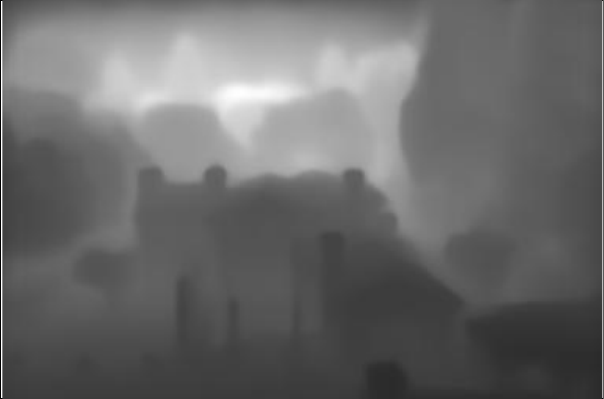} &
\includegraphics[width=0.199\linewidth]{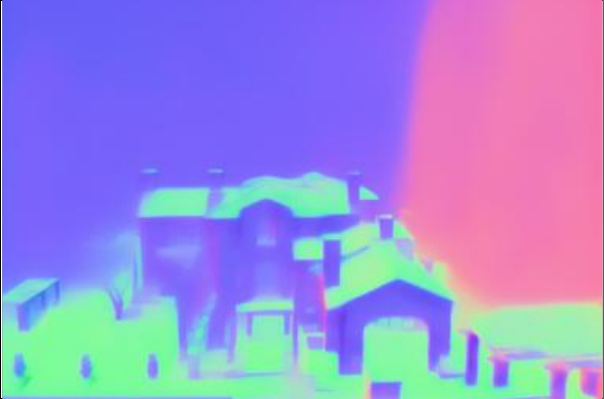} &
\includegraphics[width=0.199\linewidth]{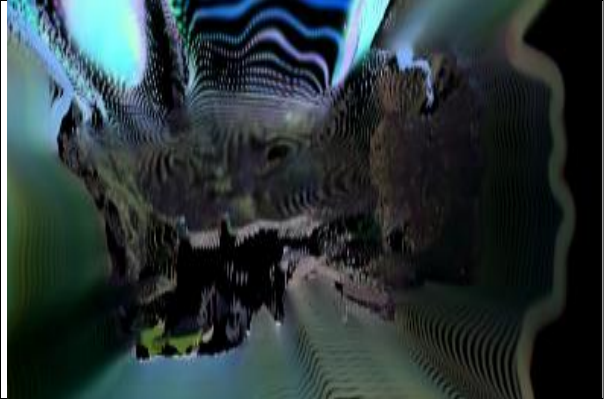} &
\includegraphics[width=0.199\linewidth]{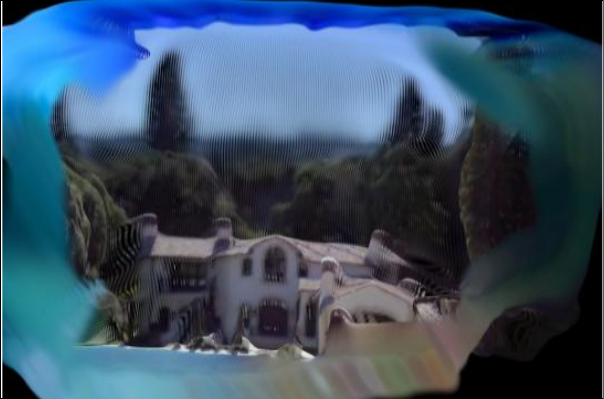} & \\
\includegraphics[width=0.199\linewidth]{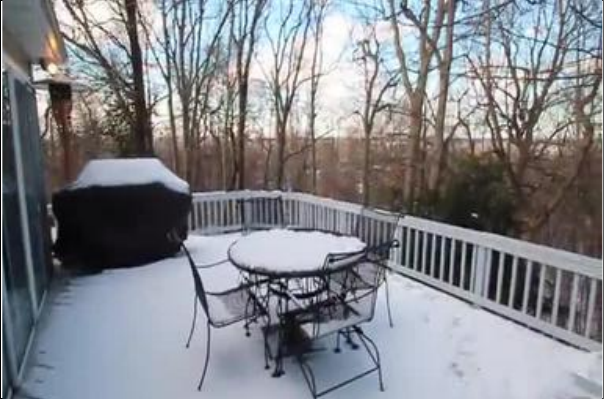} & 
\includegraphics[width=0.199\linewidth]{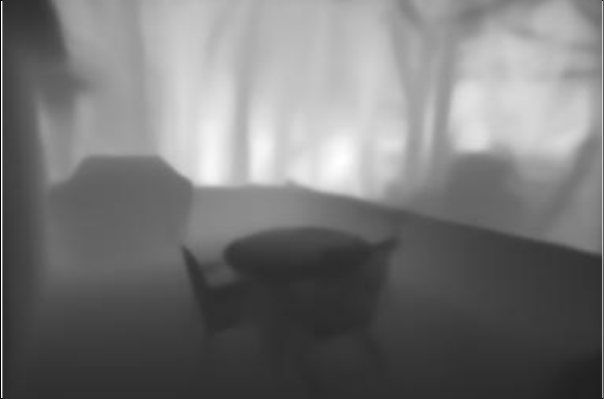} &
\includegraphics[width=0.199\linewidth]{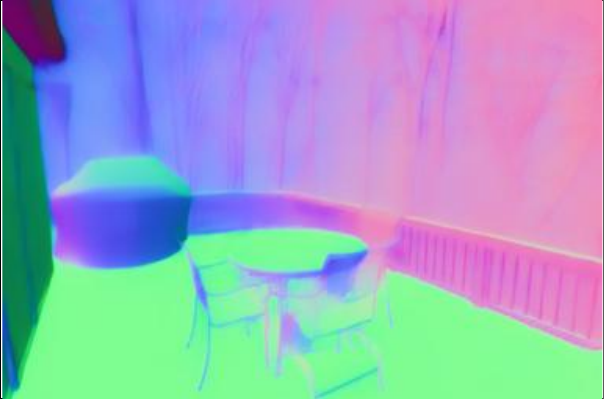} &
\includegraphics[width=0.199\linewidth]{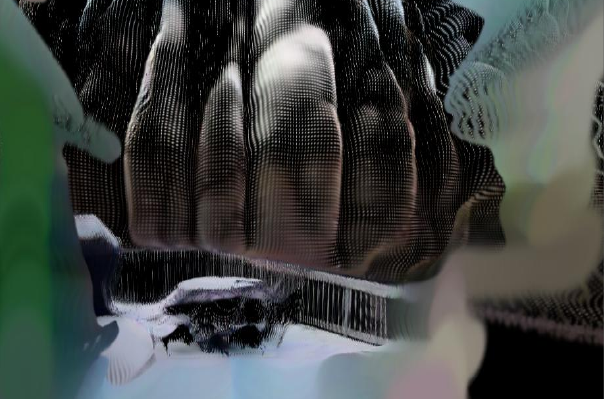} &
\includegraphics[width=0.199\linewidth]{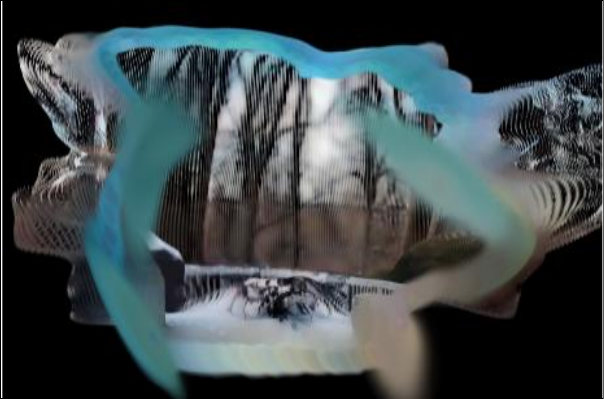} & \\
\includegraphics[width=0.199\linewidth]{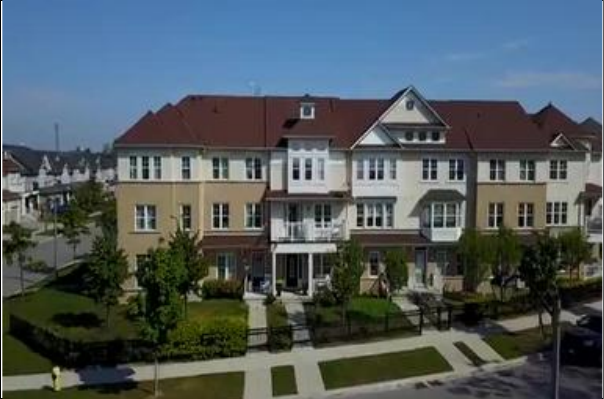} & 
\includegraphics[width=0.199\linewidth]{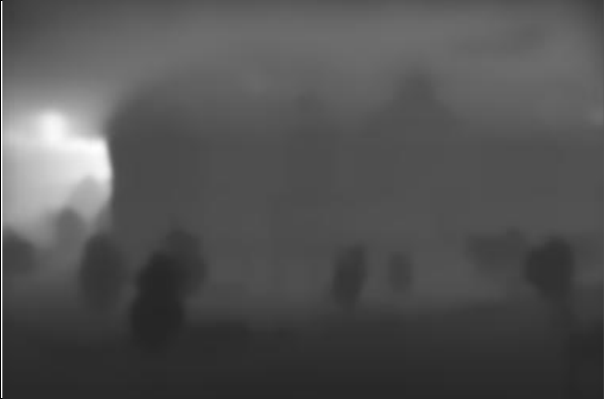} &
\includegraphics[width=0.199\linewidth]{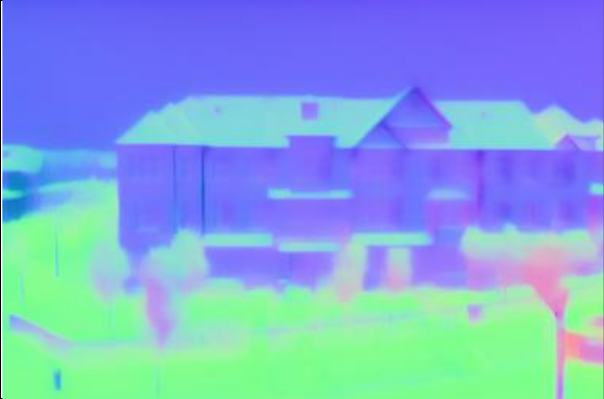} &
\includegraphics[width=0.199\linewidth]{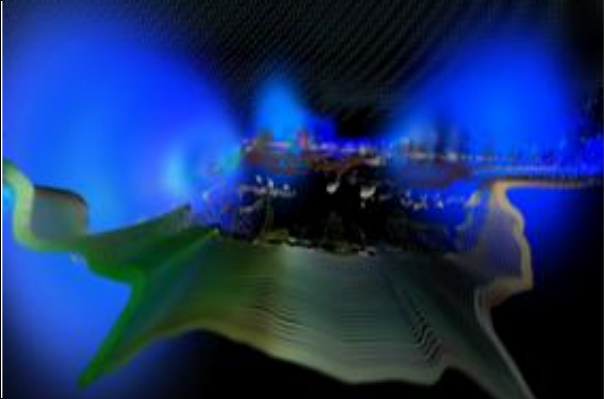} &
\includegraphics[width=0.199\linewidth]{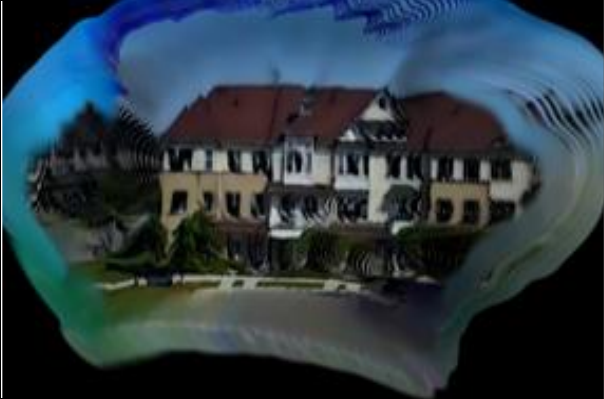} & \\
\includegraphics[width=0.199\linewidth]{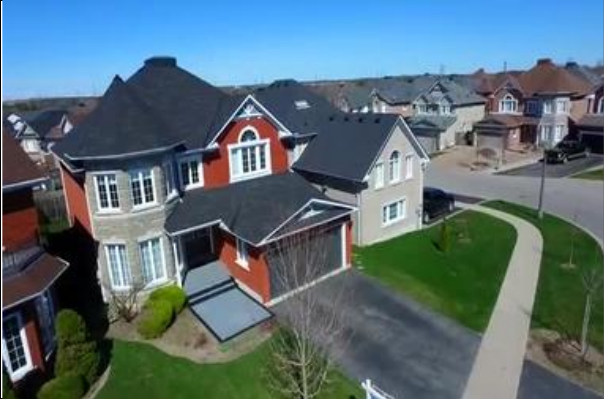} & 
\includegraphics[width=0.199\linewidth]{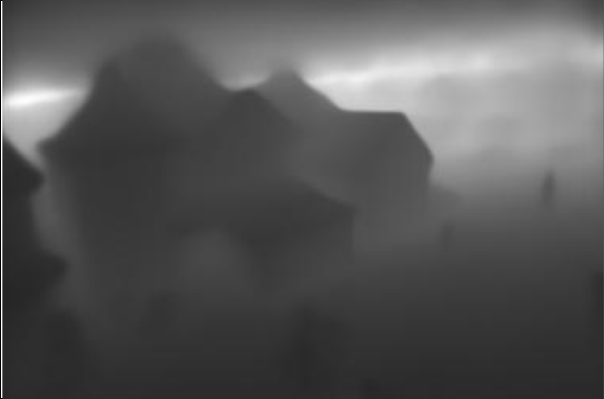} &
\includegraphics[width=0.199\linewidth]{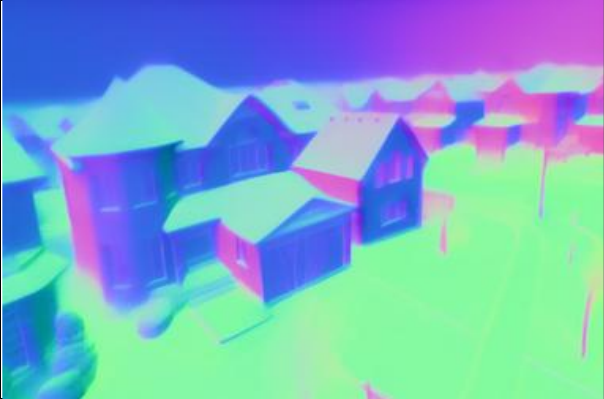} &
\includegraphics[width=0.199\linewidth]{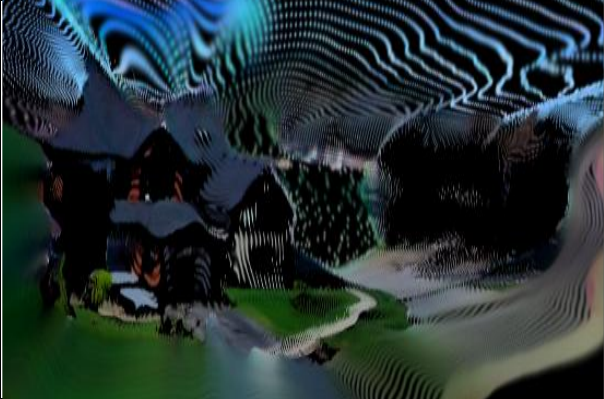} &
\includegraphics[width=0.199\linewidth]{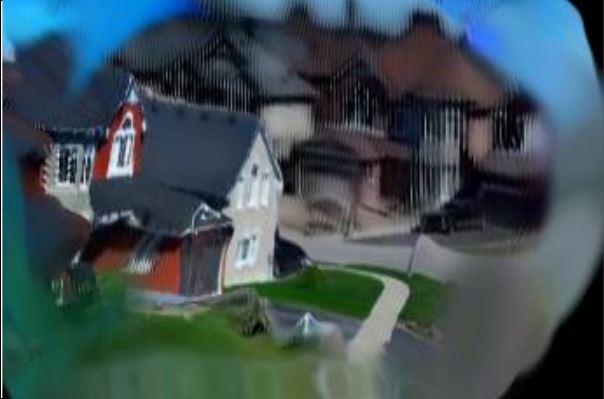} & \\
\\ \\
\small (a) Input & \small (b) Depth  & \small (c) Normal  & \small (d) Flash3D & \small (e) Ours  & \small \\
\end{tabular}
\caption{\textbf{More qualitative results.} Qualitative comparison of 3D scene reconstruction performance under varying illumination and geometric complexity. The upper panel presents indoor environments featuring intricate room layouts, while the lower panel demonstrates outdoor architectural structures with surrounding vegetation. }
\label{fig:Add qualitative results}
\vspace{-3mm}
\end{figure*}

\begin{table}
  \centering
  \small
  \begin{tabular}{@{}lcccc@{}}
    \toprule
    & \multicolumn{4}{c}{KITTI} \\
    Method & CD & \hspace{16pt}PSNR↑ & \hspace{16pt}SSIM↑ & \hspace{16pt}LPIPS↓ \\
    \cmidrule(lr){2-5}
    LDI~\cite{tulsiani2018layer} & × & \hspace{16pt}16.50 &\hspace {16pt}0.572 &\hspace{16pt} - \\
    SV-MPI~\cite{tucker2020single} & × &\hspace{16pt} 19.50 &\hspace{16pt} 0.733 &\hspace{16pt} - \\
    BTS~\cite{wimbauer2023behind} & × &\hspace{16pt} 20.10 &\hspace{16pt} 0.761 & \hspace{16pt}\underline{0.144} \\
    MINE~\cite{li2021mine} & × & \hspace{18pt}\textbf{21.90} & \hspace{18pt}\textbf{0.828} & \hspace{16pt}\textbf{0.112} \\
    Flash3D~\cite{szymanowicz2024flash3d} & \checkmark & \hspace{18pt}20.98 & \hspace{18pt}\underline{0.784} & \hspace{16pt}0.159 \\
    \bottomrule
    Ours & \checkmark & \hspace{20pt}\underline{21.24} & \hspace{18pt}0.779 &\hspace{16pt} 0.158 \\
    \bottomrule
  \end{tabular}
  \caption{Comprehensive comparison on KITTI \cite{geiger2012we} dataset. Bold values indicate better performance. In this context, cross-domain (CD) indicates that the method was not trained on the dataset being evaluated. (\textbf{Best} results are in bold, \underline{second best} underlined. )}
  \label{tab:kitti_full}
\end{table}
\vspace{-0.1cm}

As demonstrated in \autoref{fig:Add qualitative results}, we systematically evaluate Niagara against Flash3D for single-view 3D scene reconstruction under diverse illumination conditions and geometric complexity. The upper section (five rows) examines indoor environments with intricate layouts and challenging lighting, while the lower section (four rows) analyzes outdoor architectural structures surrounded by vegetation. Our method exhibits four principal advantages:

\begin{itemize}
    \item \textbf{Superior texture preservation.} Niagara consistently restores fine-grained material details across scenes. In indoor environments (e.g., Row 4, Indoor), it maintains precise transitions between wall textures, preserves sharp doorframe edges under mixed lighting, and accurately reconstructs reflective floor surfaces. For outdoor scenes (e.g., Row 2, Outdoor), the method captures layered vegetation textures on building façades, whereas Flash3D oversimplifies these details into flat regions.
    \item \textbf{Improved geometric structure reconstruction.} Niagara demonstrates enhanced recovery of spatially coherent layouts. In complex indoor settings (e.g., Row 3, Indoor), Niagara reconstructs furniture arrangements with accurate depth ordering and wall connectivity, eliminating the fragmented geometries observed in Flash3D. For outdoor structures (e.g., Row 1, Outdoor), it preserves architectural proportions such as window alignments and roof slopes, while Flash3D introduces perspective distortions.
    \item \textbf{Reduction of color bleeding and artifacts.} Niagara achieves natural color separation under challenging lighting. Indoor results (e.g., Row 5, Indoor) show distinct material boundaries between wooden furniture and painted walls, even under strong ambient light. In outdoor cases (e.g., Row 3, Outdoor), Niagara prevents vegetation hues from bleeding into adjacent stone pathways -- a common failure mode in Flash3D outputs.
    \item \textbf{Enhanced generalization to open-domain scenarios.} Niagara demonstrates robust generalization across diverse scene types. For intricate indoor spaces (e.g., multi-room layouts in Row 2, Indoor), Niagara maintains consistent scale across interconnected areas, unlike Flash3D, which struggles with occluded regions. In vegetation-heavy outdoor scenes (e.g., Row 4, Outdoor), it reconstructs overlapping foliage and architectural elements without oversimplification of natural complexity, whereas Flash3D produces flattened geometry.
\end{itemize}

These advancements originate from Niagara's novel integration of depth-normalized geometric constraints and 3D self-attention, a framework designed to resolve ambiguities in complex scene reconstruction. The depth-normal constraints enforce surface continuity through locally adaptive normal priors, effectively mitigating topological errors in occluded regions (e.g., overlapping foliage, multi-room junctions). Concurrently, the 3D self-attention mechanism learns long-range structural dependencies across scales, dynamically harmonizing geometric coherence. This dual strategy overcomes the rigidity of prior-based methods like Flash3D, which rely on static scene assumptions and lack mechanisms to refine geometry and appearance.

\section{KITTI Experiments}  

We present experimental results on the KITTI \cite{geiger2012we} dataset, demonstrating our model's enhanced capability for outdoor scene reconstruction compared to Flash3D \cite{szymanowicz2024flash3d}. Technical constraints prevented the direct reproduction of metrics reported in the original Flash3D publication, though we remain in communication with the authors to resolve methodological discrepancies.

As shown in \autoref{tab:kitti_full}, our approach outperforms Flash3D under equivalent configurations. However, both methods exhibit deviations from standard evaluation protocols, particularly in Flash3D's open-source implementation (setting an in \autoref{tab:kitti_full}). We attribute performance variations primarily to differences in KITTI dataset preprocessing methodologies.

While none of the results achieve optimal performance metrics, this analysis focuses on comparative configuration impacts rather than absolute performance evaluation.


\begin{IEEEbiography}[{\includegraphics[width=0.9in,height=1.1in,clip,keepaspectratio]{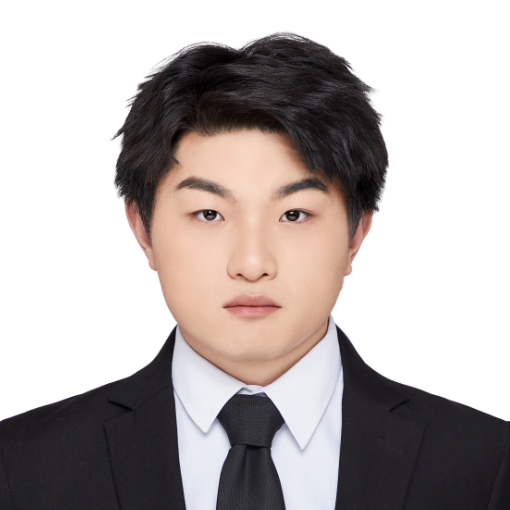}}]{Xianzu Wu}
obtained his B.E. in Data Science and Big Data Technology from Yangtze University in 2024. He has conducted research at Westlake University and SUNY Buffalo on 3D scene reconstruction and point cloud completion. His interests include computer vision and multi-modal modeling. He has published in CVPR. More at: \url{https://xianzuwu.github.io}.
\end{IEEEbiography}

\vspace{-2mm}

\begin{IEEEbiography}[{\includegraphics[width=0.9in,height=1.1in,clip,keepaspectratio]{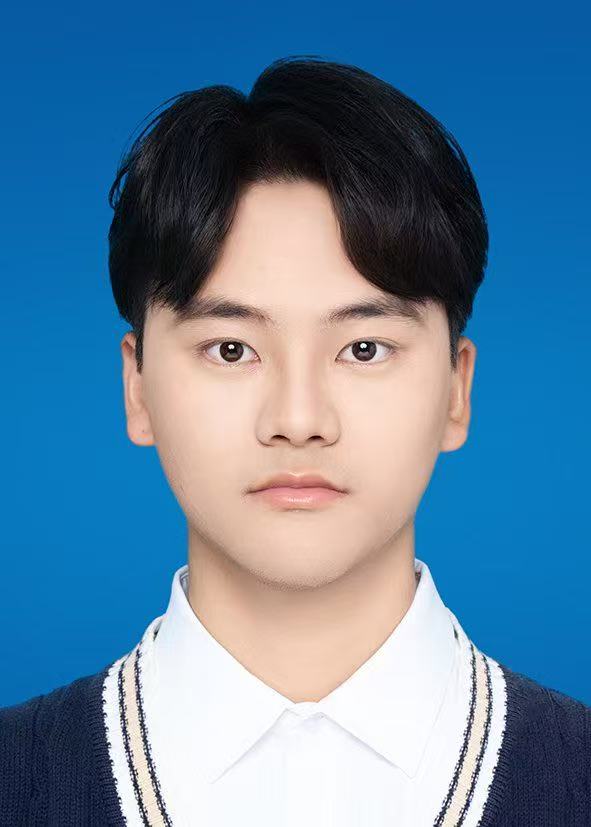}}]{Zhenxin Ai} is currently pursuing his B.E. in Artificial Intelligence at Jiangxi University of Science and Technology, with a visiting stint at Westlake University. His research focuses on computer vision and large language models, specifically spanning lightweight salient object detection, monocular 3D scene reconstruction, and LLM. He has conducted research at Jiangxi Province Key Laboratory of Multidimensional Intelligent Perception and Control of china (supervised by Prof. Luo Huilan), Encode Lab of Westlake University (supervised by Prof. Huan Wang). His work has led to a first-author publication in IEEE TGRS.  More at: \url{https://ai-kunkun.github.io/}.
\end{IEEEbiography}

\vspace{-2mm}

\begin{IEEEbiography}[{\includegraphics[width=0.9in,height=1.1in,clip,keepaspectratio]{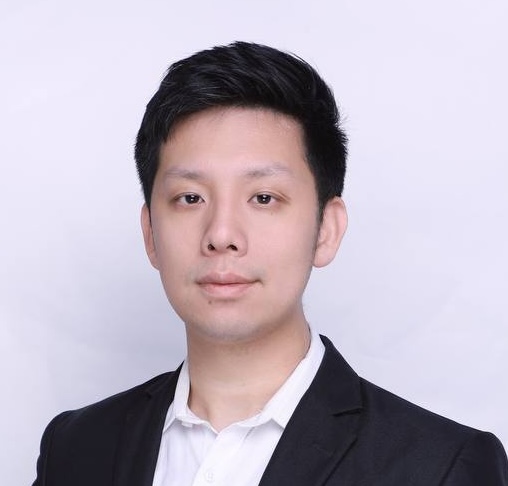}}]{Harry Yang}
 is an Assistant Professor with the Division of Art and Machine Creativity at HKUST. Previously, he was a Senior Scientist at Meta AI. He received his Ph.D. from the University of Southern California and his B.S. from the University of Science and Technology of China. His research focuses on video and multimodal generation.
\end{IEEEbiography}
\vspace{-3mm}

\begin{IEEEbiography}[{\includegraphics[width=0.9in,height=1.1in,clip,keepaspectratio]{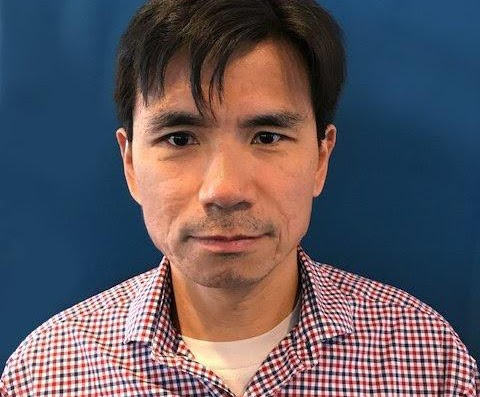}}]{Sernam Lim}
 concentrates on Computer Vision and
the field of AI. He pursued and earned a PhD at the
University of Maryland, College Park, in 2005. SerNam spent a decade at GE Research working on
different areas of Computer Vision including video
recognition, 3D reconstruction, representation, and
visual matching. At GE Research, Ser-Nam started
as a Research Scientist, then took a role as the
Computer Vision Lab Manager, and finished his
career there as a Senior Principal and Director. SerNam then took a role managing multiple AI teams
at Meta that conduct research in Computer Vision, NLP and other areas
of AI with a focus on applying and scaling to huge amounts of data on
the Meta platform. Ser-Nam’s work has accomplished impactful production
outcomes in ensuring safety in the Aviation and Power industry as well as
detecting misinformation and other integrity issues on the Meta platform. At
the end of his career at Meta, Ser-Nam leads projects focused on AI for
user to content recommendations, as well as search engines that include the
intersection of Large Language Models (LLM) and Computer Vision. After
Meta, Ser-Nam joined the Computer Science faculty at University of Central
Florida (UCF) as a tenured Associate Professor. His group at UCF conducts
research in image and video generation, AI for Augmented Reality, visual language representation and understanding and other major topics in AI. SerNam has published over 200 peer-refereed papers, with more than half in top AI venues.
\end{IEEEbiography}
\vspace{-3mm}

\begin{IEEEbiography}[{\includegraphics[width=0.9in,height=1.1in,clip,keepaspectratio]{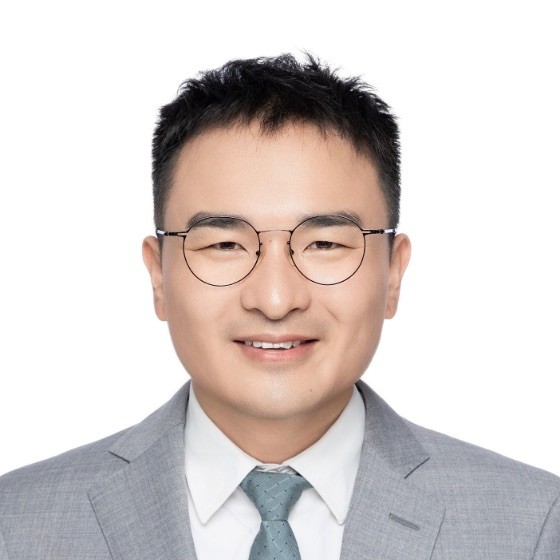}}]{Jun Liu}
(Senior Member, IEEE) is a Professor at School of Computing and Communications in Lancaster University. He got the PhD degree from Nanyang Technological University in 2019. He was with Singapore University of Technology and Design from 2019 to 2024. He obtained the best paper awards from PREMIA in 2016 and 2019, the Best Doctoral Thesis Award from EEE at NTU in 2020, and the IEEE VSPC Rising Star Honorable Mention Award in 2024. He is a Senior Area Editor of TIP, and an Associate Editor of TCSVT, TII, TBIOM, CSUR, and PR. His research interests include computer vision, machine learning and digital health.
\end{IEEEbiography}

\vspace{-3mm}

\begin{IEEEbiography}[{\includegraphics[width=0.9in,height=1.1in,clip,keepaspectratio]{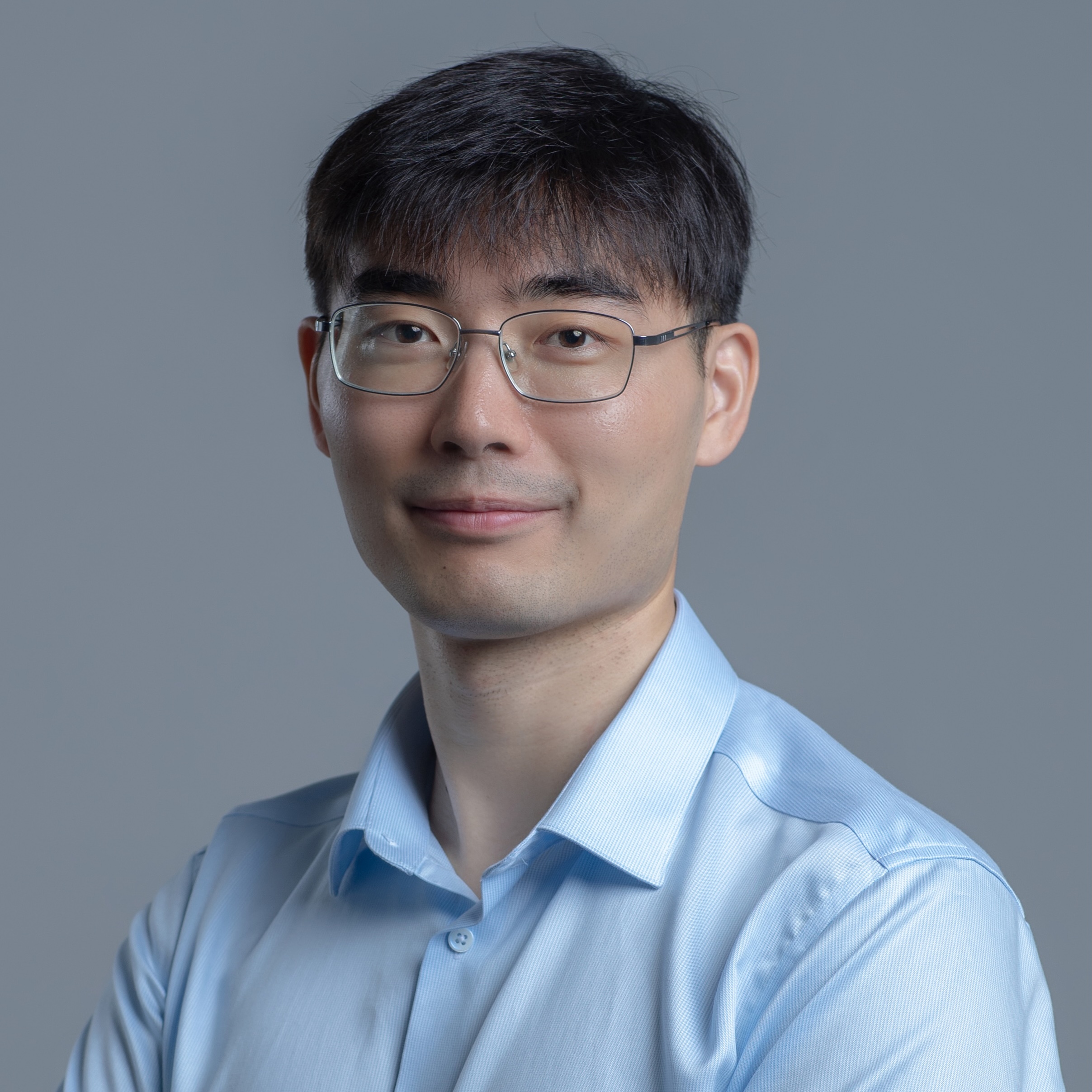}}]{Huan Wang}
(Member, IEEE) is currently a Tenure-Track Assistant Professor at Department of Artificial Intelligence, Westlake University (Hangzhou, China), leading ENCODE (Efficient Neural Computing and Design) Lab. He earned his PhD degree (2024) at Northeastern University (Boston, USA), advised by Prof. Yun Raymond Fu (FIEEE/NAI/AAAS, MAE). Before that, he obtained his M.S. degree (2019) and B.E. degree (2016) at Zhejiang University (Hangzhou, China). He works on Efficient AI, machine learning, and computer vision, with over 30 papers in the field of computer vision and machine learning. Most of them appear in top-tier venues (e.g., CVPR/ICCV/ECCV/NeurIPS/ICLR/TPAMI/TIP). He also works closely with industrial researchers from Snap / Salesforce / Amazon / Alibaba, etc. See more information at: \url{https://huanwang.tech}.
\end{IEEEbiography}

\end{document}